\documentclass[%
 reprint,%
superscriptaddress,
 amssymb, amsmath,%
 aps,pre,%
]{revtex4-1}
\usepackage{docs}%
\usepackage{dcolumn}
\usepackage{graphicx}
\usepackage{xcolor}
\usepackage{amsfonts}
\usepackage{amsmath}
\usepackage{amssymb}
\usepackage{amsthm}
\usepackage{bm}
\usepackage[breaklinks=true]{hyperref}
\hypersetup{
  colorlinks   = true, %Colours links instead of ugly boxes
  urlcolor     = blue, %Colour for external hyperlinks
  linkcolor    = blue, %Colour of internal links
  citecolor   = blue %Colour of citations
}

\date{\today}

\begin{document}

\title{Mixing and transport from combined stretching-and-folding and cutting-and-shuffling}

\author{Lachlan~D. Smith}
 \email{lachlan.smith@northwestern.edu}
   \affiliation{ 
Department of Chemical and Biological Engineering, Northwestern University, Evanston, IL 60208, USA%\\This line break forced with \textbackslash\textbackslash
}%
\author{Paul~B. Umbanhowar}
\affiliation{Department of Mechanical Engineering, Northwestern University, Evanston, IL 60208, USA}
\author{Julio~M. Ottino}
 \affiliation{ 
Department of Chemical and Biological Engineering, Northwestern University, Evanston, IL 60208, USA%\\This line break forced with \textbackslash\textbackslash
}%
\affiliation{Department of Mechanical Engineering, Northwestern University, Evanston, IL 60208, USA}
\affiliation{The Northwestern Institute on Complex Systems (NICO), Northwestern University, Evanston, IL 60208, USA}
\author{Richard~M. Lueptow}
 \email{r-lueptow@northwestern.edu}
 \affiliation{ 
Department of Chemical and Biological Engineering, Northwestern University, Evanston, IL 60208, USA%\\This line break forced with \textbackslash\textbackslash
}%
\affiliation{Department of Mechanical Engineering, Northwestern University, Evanston, IL 60208, USA}
\affiliation{The Northwestern Institute on Complex Systems (NICO), Northwestern University, Evanston, IL 60208, USA}

\begin{abstract}  
While structures and bifurcations controlling tracer particle transport and mixing have been studied extensively for systems with only stretching-and-folding, and to a lesser extent for systems with only cutting-and-shuffling, few studies have considered systems with a combination of both. We demonstrate two bifurcations for non-mixing islands associated with elliptic periodic points that only occur in systems with combined cutting-and-shuffling and stretching-and-folding, using as an example a map approximating biaxial rotation of a less-than-half-full spherical granular tumbler. First, we characterize a bifurcation of elliptic island containment, from containment by manifolds associated with hyperbolic periodic points to containment by cutting line tangency. As a result, the maximum size of the non-mixing region occurs when the island is at the bifurcation point. We also demonstrate a bifurcation where periodic points are annihilated by the cutting-and-shuffling action. Chains of elliptic and hyperbolic periodic points that arise when invariant tori surrounding an elliptic point break up [according to Kolmogorov-Arnold-Moser (KAM) theory] can annihilate when they meet a cutting line. Consequently, the Poincar\'{e} index (a topological invariant of smooth systems) is not preserved. Copyright \copyright 2017 American Physical Society. Published in Physical Review E \textbf{96}, 042213 (2017).

\noindent DOI: \href{https://doi.org/10.1103/PhysRevE.96.042213}{10.1103/PhysRevE.96.042213}
\end{abstract}

\maketitle

\section{Introduction} \label{sec:intro}

While the dynamics governing passive tracer transport and mixing are well understood for classical smooth systems \cite{Ottino, Wiggins+Ottino}, where mixing is created by stretching-and-folding, and recent advances have improved the understanding of mixing and transport in piecewise isometries (PWIs) in one dimension \cite{Ashwin2002, Avila2007, Novak2009, Christov2011, Krotter2012, Yu2016}, two-dimensional (2D) planar \cite{Haller1981, Ashwin1997, Gutkin1997, Goetz2003, Sturman2012, Hughes2012, Hughes2013}, and 2D curvilinear geometries \cite{Scott2001, Scott2003, Juarez2010, Juarez2012, Park2016, Park2017, Smith2017BSTresonances}, where only cutting-and-shuffling is present, few studies have considered mixing and transport in systems with combined stretching-and-folding and cutting-and-shuffling actions. A multitude of natural and engineered systems exhibit combined stretching-and-folding and cutting-and-shuffling actions, including granular flows \cite{Juarez2012, Zaman2017}, fluid flows with valves \cite{Jones1988, Hertzsch2007, Beuf2010, Smith2016discdef, Smith2017LSID}, yield stress materials \cite{Boujlel2016, Louzguine2012, Olmsted2008, Brown2014}, geological faults \cite{Boyer1982}, and splitting and recombining droplets \cite{Hunt2008, Schultz2014}. It has been shown that in systems with both stretching-and-folding and cutting-and-shuffling, the cutting action can create gaps in structures that represent transport barriers in classical smooth systems. Tracer particles are able to pass through the gaps, enabling greater topological freedom for transport \cite{Smith2016discdef}. For three-dimensional systems with both stretching-and-folding and cutting-and-shuffling, the cutting action can provide a mechanism for 3D transport (i.e.\ trajectories are not confined to 1D curves or 2D surfaces) \cite{Smith2017LSID}. Counter-intuitively, the addition of cutting to a smooth system can decrease the rate of mixing due to the creation of new structures termed pseudo-periodic points \cite{Smith2017CSSrate}.

Here we consider mixing and transport generated by a map approximating granular flow in a less-than-half-full biaxial spherical tumbler (BST) in which each iteration consists of a rotation of the tumbler about the $z$-axis by $\theta_z$, followed by rotation about the $x$-axis by $\theta_x$. We use the infinitely thin flowing layer approximation \cite{Sturman2008}. Remarkably, this simplified model, which is valid on 2D spherical shells, has been shown to accurately predict transport structures and non-mixing regions observed in experiments \cite{Zaman2017}, even though 3D effects, diffusion in the flowing layer, and shear in the flowing layer are neglected. So far, however, only the half-full spherical tumbler scenario has been considered, where the approximating map is a PWI and no stretching-and-folding occurs. Here we relax that constraint. When the tumbler is less than half-full we show that each flowing layer crossing yields a shear, and by alternating rotation axes (between $z$- and $x$-axis rotation), stretching-and-folding can be produced alongside cutting-and-shuffling. By studying advective transport in the BST flow, we aim to improve the understanding of mixing and transport in systems with combined cutting-and-shuffling and stretching-and-folding motions. This could lead to improvements in industrial mixers with more complicated geometries \cite{Lemieux2007, Doucet2008, Alizadeh2013, Golshan2017, Ottino2008, Vargas2008, Jiang2011, Simons2016, Hassanpour2011}, especially since many granular mixers would benefit from a mixing mechanism that operates at larger length scales than those associated with diffusion.

\begin{figure}[tbp]
\centering
\includegraphics[width=\columnwidth]{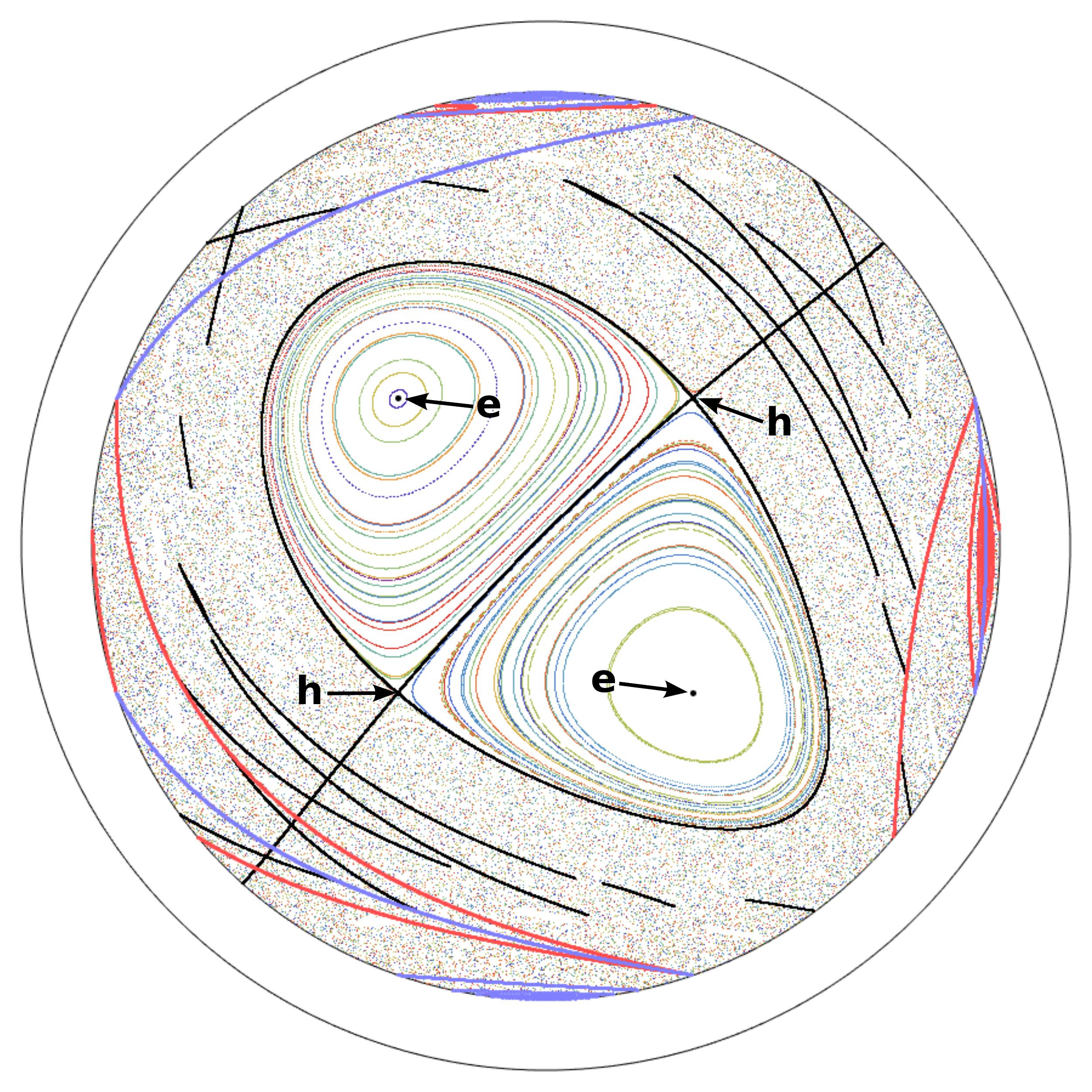}
\caption{Poincar\'{e} section for the BST map with $(\theta_z,\theta_x,h)= (2.045,2.045,0.5)$, where $h$ is the orthogonal distance from the rotation axis to the center of the flowing layer (bottom view, see \S\ref{sec:BST_map} for full details of the BST map). Different colors represent trajectories of different points. Two period-1 islands with elliptic periodic points at their centers (black points indicated by `e') are contained by the stable and unstable manifolds (thin black curves) associated with period-1 hyperbolic points (marked by `h'). Cutting lines, where material is cut after a single iteration of the BST map, are indicated by the thick blue and red (gray) curves. Thick blue curves indicate the cuts produced by the $z$-axis rotation, and thick red curves indicate the cuts produced by $x$-axis rotation. The outer circle has radius equal to 1, representing the boundary of the unit hemispherical shell, and the inner circle has radius $\sqrt{1-h^2}=\sqrt{3}/2$, representing the boundary of the map domain.}
\label{fig:containment_psection_1}
\end{figure}

The key question is how classical structures controlled by stretching-and-folding interact with cutting-and-shuffling. In particular, how do non-mixing islands associated with elliptic periodic points interact with cutting lines (where material is cut after an iteration of the map)? For instance, Fig.~\ref{fig:containment_psection_1} shows a typical Poincar\'{e} section for the BST map, with two period-1 islands contained by the stable and unstable manifolds (thin black curves) associated with a pair of period-1 hyperbolic points (marked by `h'). As the system parameters vary, the periodic points, manifolds and cutting lines [thick red and blue (gray) curves] move, and the period-1 islands change in size. We show that at critical parameter values the islands meet the cutting lines, leading to a transition from containment of the islands by stable and unstable manifolds (i.e.\ classical smooth phenomena, as occurs in Fig.~\ref{fig:containment_psection_1}) to containment by cutting line tangency (i.e.\ the manifolds are cut, and the island perimeter is tangent to a cutting line, like islands in pure PWIs). This containment bifurcation is important for transport and mixing because when cutting is changed to a localized shear (e.g.,\ arising from a finite thickness flowing layer in a granular tumbler compared to the infinitesimally thin limit considered here), cutting lines become `thickened', and elliptic islands tangent to cutting lines shrink as the cutting lines thicken. On the other hand, thickening of the cutting lines does not affect an elliptic island contained by manifolds away from cutting lines, making them more robust and more likely to influence mixing and transport in an experimental granular tumbler \cite{Zaman2017}.

Since elliptic islands can never cross a cutting line (otherwise they would be cut), restricting the domain of a combined stretching-and-folding and cutting-and-shuffling map to an island, or a set of islands, eliminates the cutting-and-shuffling action, and the map is a diffeomorphism (smooth with a smooth inverse). Consequently, intra-island transport is governed by Kolmogorov-Arnold-Moser (KAM) theory, and, as predicted by the Poincar\'{e}--Birkhoff theorem, invariant tori surrounding elliptic periodic points break up into alternating chains of elliptic and hyperbolic periodic points within islands \cite{Ottino}. However, novel phenomena occur when the chains of elliptic and hyperbolic points meet a cutting line. We show that the chains of elliptic and hyperbolic points can annihilate, an impossibility for classical smooth systems, where the Poincar\'{e} index must be conserved \cite{Katok}. In essence, the cutting lines act as domain boundaries, separating regions of distinct smooth flow, although tracer particles are able to move between the separated regions.

\section{The biaxial spherical tumbler map} \label{sec:BST_map}

Before discussing two bifurcations unique to systems with combined stretching-and-folding and cutting-and-shuffling in \S\ref{sec:bifurcations}, we first describe the BST flow and map. In particular, it is necessary to find and classify the periodic points of the BST map.

\begin{figure}[tbp]
\centering
\includegraphics[width=\columnwidth]{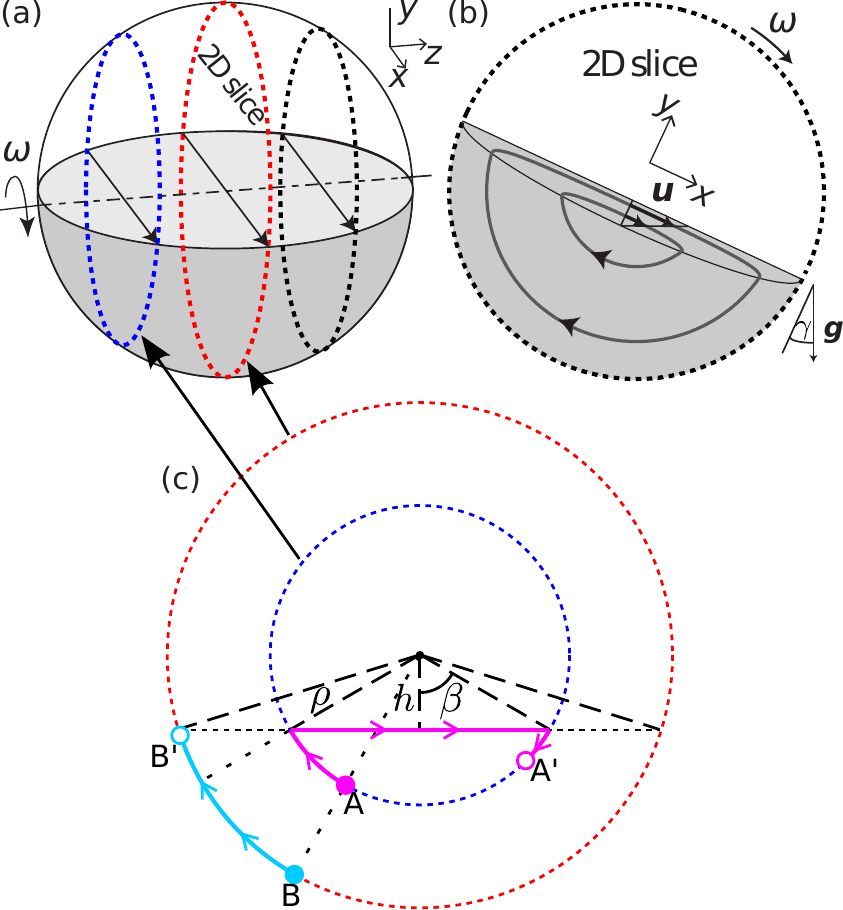}
\caption{Particle motion in the BST flow. (a)~For single axis rotation of granular material the flow can be approximated as a set of 2D slices, as the axial velocity component is small \cite{Zaman2013}. Note that $y$ is normal to the flowing layer surface. (b)~The flow in one 2D slice when the tumbler is half full. Tracer particles in the bulk experience solid body rotation until they reach the flowing layer at some position $(x,y)$. Particles rapidly flow down the flowing layer, exiting at the reflected position $(-x,y)$. (c)~Flow on the perimeter of two 2D slices when the tumbler is less than half full (i.e.,\ $h>0$) and the flowing layer is infinitely thin [particles instantaneously reflect from $(x,y)$ to $(-x,y)$ when they reach the flowing layer]. Because the subtended angle, $2\beta = 2\arccos(h/\rho)$, from the entry to the exit of the flowing layer is smaller for the circle closer to the pole of the tumbler (inner blue dashed circle), the pink particle starting at A crosses the flowing layer sooner than the cyan particle starting at B, even though they have the same initial longitude. Adapted with permission from Zaman \emph{et al.}, Phys. Rev. E, {\bf 88}(1), 012208 (2013) \cite{Zaman2013}. \copyright 2013 American Physical Society.}
\label{fig:BST_schematic}
\end{figure}

Following the continuum model introduced by Sturman \emph{et al.} \cite{Sturman2008}, flow in a less-than-half-full spherical tumbler in the continuous flow regime can be approximated as a nested set of 2D flows, with each 2D flow having two disjoint regions: the bulk, where particles act as a solid, experiencing only solid body rotation, and the thin flowing layer at the free surface, where particles flow like a fluid, as shown in Fig.~\ref{fig:BST_schematic}(a,b) for the half-full case. The flow can be further approximated by considering the vanishing flowing layer thickness limit. In this limit, tracer particles instantaneously cross the flowing layer, meaning particles are reflected across the center of the flowing layer. Each rotation of the tumbler maps the domain $S=\{\bm{x}=(x,y,z): \, |\!| \bm{x} |\!| \leq R, \, y\leq -h\}$ to itself, where $y$ is normal to the flowing layer surface, consistent with the convention for 2D tumbler flow, $R$ is the radius of the sphere, and $R - h$ is the fill depth, i.e.\ $h$ is the orthogonal distance from the rotation axis to the center of the flowing layer, as shown in Fig.~\ref{fig:BST_schematic}(c). For a clockwise $z$-axis rotation of the tumbler through an angle $\theta_z$, the map can be expressed as
\begin{equation}
M_{\theta_z}^z (\bm{x}) = R_\gamma^z (\bm{x}),
\end{equation}
where $R_\gamma^z$ denotes rotation by $\gamma$ about the $z$-axis, $\gamma = - \theta_z + 2 m \beta$, $m$ denotes the number of times the particle crosses the flowing layer, given by
\begin{equation}
m = -\left\lfloor \frac{ \arg(x+iy) -\theta_z + \beta + \pi/2 } {2 \beta } \right\rfloor,
\end{equation} 
where $\lfloor a \rfloor$ is the floor of $a$ (i.e.\ the greatest integer less than $a$), $\beta = \arccos (h/\rho)$ is half the subtended angle of the flowing layer with respect to the rotation axis, and $\rho = \sqrt{x^2+y^2}$, as shown in Fig.~\ref{fig:BST_schematic}. The map $M_{\theta_x}^x$ corresponding to anti-clockwise rotation of the tumbler about the $x$-axis by $\theta_x$ can be derived in the same way, or can be written in terms of $z$-axis rotation as
\begin{equation}
M_{\theta_x}^x (\bm{x}) = R_{-\pi/2}^y M_{\theta_x}^z R_{\pi/2}^y (\bm{x}). 
\end{equation}

The full BST map consists of rotation $\theta_z$ about the $z$-axis followed by rotation $\theta_x$ about the $x$-axis, or $M_{\theta_z,\theta_x}=M_{\theta_x}^x M_{\theta_z}^z$. As both the $z$- and $x$-axis rotations preserve the distance of a tracer particle from the origin, $\sqrt{x^2 + y^2 + z^2}$, tracer particles are confined to spherical shells. Furthermore, as long as the radius $R$ and distance $h$ have constant ratio, then the intrasurface transport of tracer particles is identical up to a scale factor (see Appendix~\ref{sec:LR_equiv}). Therefore, we restrict our attention to transport on the unit hemispherical shell with $R=1$. The only control parameters are $h$ and the rotation angles $\theta_z,\theta_x$, and we refer to a trio $(\theta_z,\theta_x,h)$ as a protocol.

Previous studies of the BST map focused exclusively on the half-full case, $h=0$, which is a piecewise isometry (PWI), thus performing only cutting-and-shuffling actions \cite{Juarez2010, Juarez2012, Sturman2012, Park2016, Park2017, Zaman2017, Smith2017BSTresonances}. In contrast, when $h > 0$, stretching-and-folding actions can occur due to the presence of azimuthal shear. To understand why, consider the trajectories of two tracer particles on the unit hemispherical shell with the same initial longitude under single axis rotation with $h > 0$ [Fig.~\ref{fig:BST_schematic}(c)]. The subtended angle of the flowing layer with respect to the rotation axis, $2\beta$, is smaller for the tracer particle closer to a pole (pink particle starting at A on the inner blue dashed circle). Therefore, the pink particle reaches and crosses the flowing layer before the cyan particle starting at B, and shear results.

\begin{figure*}[tbp]
\centering
\includegraphics[width=\textwidth]{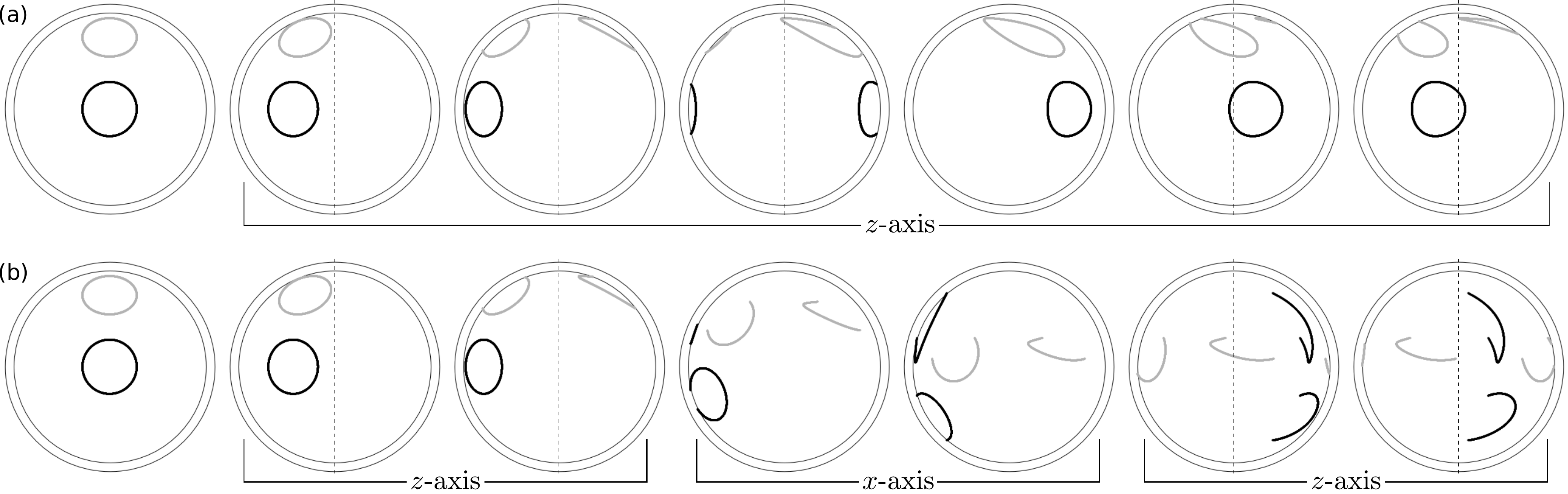}
\caption{Tracking two initially circular sets of tracer particles under the BST map (bottom view) with $h=0.4$. The outer circle has radius equal to $1$, representing the boundary of the unit hemispherical shell, and the inner circle has radius $\sqrt{1-h^2} \approx 0.917$, representing the boundary of the map domain, $\partial S$, due to $h\neq 0$. The rotation axis is indicated by the dashed line. (a)~$z$-axis rotation through $4\pi/5 $. (b)~BST map with $\theta_z = \theta_x = 4\pi/15$.}
\label{fig:circle_tracking}
\end{figure*}

To further demonstrate the effect of this shear, consider the trajectories of the initially circular black and gray sets of particles in Fig.~\ref{fig:circle_tracking}(a), for which the rotation is $\theta_z = 144^\circ$ with no rotation about the $x$-axis. Since $\beta$ decreases toward the poles, the circles are sheared each time they cross the flowing layer, with zero shear at the equator, and increasing shear nearer the poles. Furthermore, $\beta$ being smaller nearer the poles means particles cross the flowing layer more frequently (e.g.,\ the black circle near the equator crosses the flowing layer once, whereas portions of the gray circle near the pole cross the flowing layer twice), and the increased shearing is exacerbated. The shear produced by single axis rotation leads to increases in the lengths of the black and gray curves, but the growth is sub-exponential as no stretching-and-folding results. In contrast, in the biaxial flow shear is generated in multiple directions, and stretching-and-folding can be produced. Furthermore, alternating the rotation axis also results in cutting-and-shuffling like the half-full case \cite{Sturman2008}. For example, the initially circular sets of tracer particles in Fig.~\ref{fig:circle_tracking}(b) are cut if the rotation axis is changed when they have only partially crossed the flowing layer, and they are sheared during each flowing layer crossing. 

We can determine where cuts will occur by finding the points that are mapped onto the boundary of the domain, $\partial S$, after the $z$-axis rotation and after both the $z$- and $x$-axis rotations in Fig.~\ref{fig:circle_tracking}(b). We refer to these curves as ``cutting lines,'' and they are demonstrated by the thick red and blue (gray) curves in Fig.~\ref{fig:containment_psection_1} for the protocol $(\theta_z,\theta_x,h)=(2.045,2.045,0.5)$. A cluster of particles that straddles a cutting line will be cut and rearranged in the next iteration of the BST map. Cutting lines are important in PWIs since the size of each non-mixing island is equal to the distance from the elliptic periodic point at the center of the island to the nearest cutting line \cite{Smith2017BSTresonances}.

Based on Fig.~\ref{fig:circle_tracking}(b) it is evident that the less than half-full BST map is not a PWI nor a classical smooth dynamical system. The map is smooth on each of the pieces separated by the cutting lines, but it is also not a so-called ``piecewise-smooth map'' \cite{Wojtkowski1982, Bullett1986, Banerjee1999, Bernardo2008, Makarenkov2012}, because piecewise smooth maps are continuous at the borders where pieces meet (i.e. no cutting-and-shuffling occurs). In \S\ref{sec:bifurcations} we show that it produces a combination of classical dynamics (KAM theory, manifolds as transport barriers, etc.) and PWI dynamics (island size governed by distance to cutting lines, periodic point annihilation, etc.), and explore the interplay and interaction between the two dynamics. Before proceeding, though, it is useful to describe the nature of period-1 points of the BST map, whose associated structures play a central role in later analysis.

\subsection{Period-1 points} \label{sec:period_1_points}

We focus on the period-1 points of the BST map because they can be identified and studied analytically. Furthermore, their properties (e.g.,\ shape, interaction with cutting lines, and bifurcations) are representative of other periodic points with higher periodicity.

\begin{figure}[tbp]
\centering
\includegraphics[width=\columnwidth]{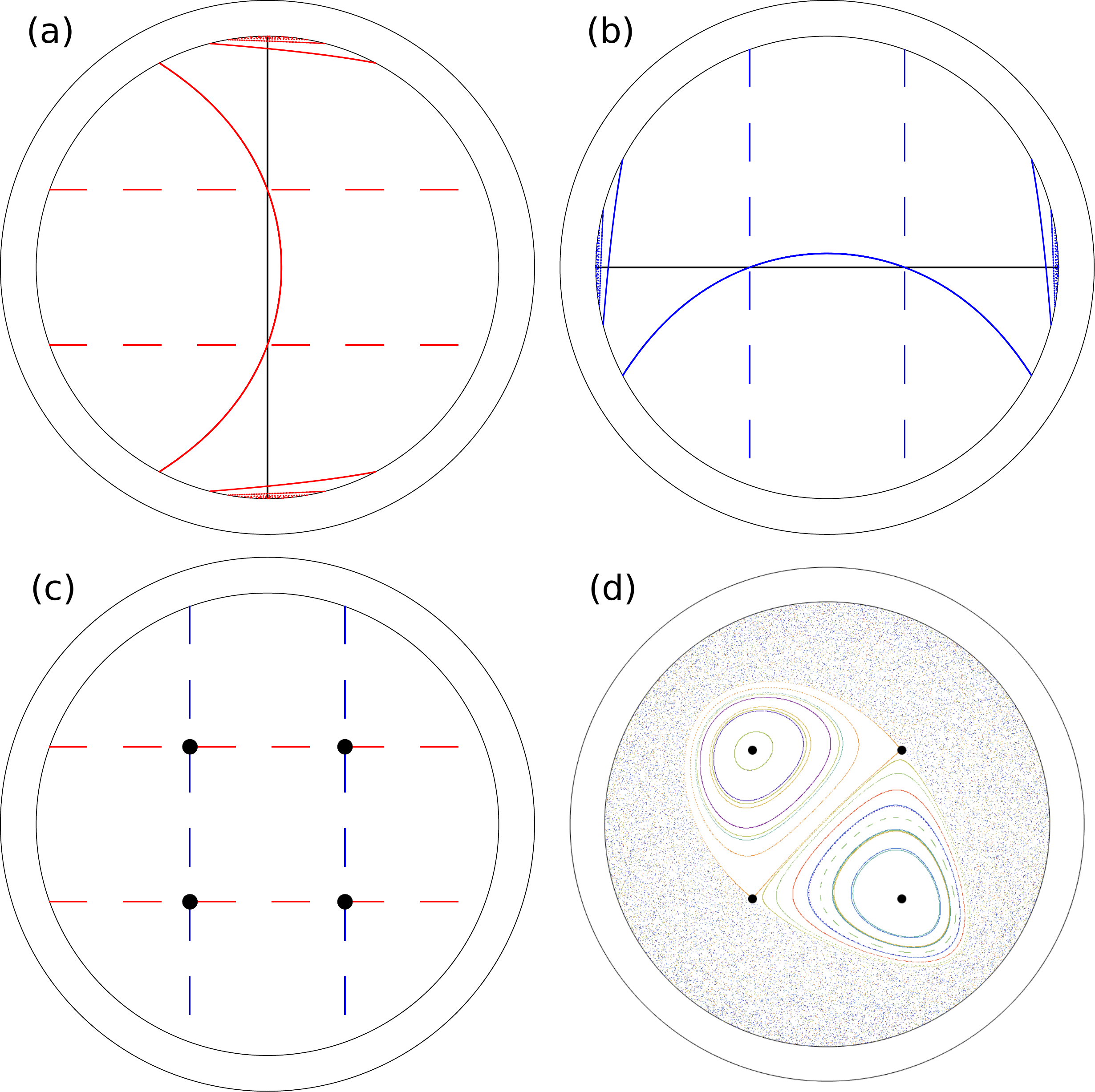}
\caption{Locating period-1 points for $(\theta_z,\theta_x,h)=(13\pi/20,13\pi/20,0.5)$ (bottom view). The outer circle has unit radius, and the inner circle with radius $\sqrt{1-h^2}\approx 0.866$ is the boundary of the domain, $\partial S$. (a)~Image (solid red) of the initial tracer curve $x=0,y^2+z^2=1$ (black vertical curve) on the unit hemispherical shell under $M_{\theta_z}^z$. All points on the dashed red curves of constant latitude (with respect to the $z$-axis rotation) return to their initial position after the $z$-axis rotation. (b)~Same as (a) except for the $x$-axis rotation, mapping the horizontal curve $z=0,x^2+y^2=1$ (black). Points on the dashed blue curves of constant latitude (with respect to the $x$-axis rotation) return to their initial position after the $x$-axis rotation. (c)~The four points of intersection of the dashed lines in (a,b) are period-1 points, as they return to their initial position after completing both the $z$- and $x$-axis rotations. (d)~Corresponding Poincar\'{e} section, where different colors show the orbits of 50 randomly seeded initial tracer particles.}
\label{fig:P1_schematic}
\end{figure}

Since the $z$-axis rotation only changes the $x$ and $y$ coordinates of a tracer particle, and the $x$-axis rotation only changes the $y$ and $z$ coordinates of a particle, period-1 points of the BST map must return to their starting position after each rotation phase, i.e.\ $M_{\theta_z}^z (\bm{x}) = M_{\theta_x}^x (\bm{x}) = \bm{x}$. Considering first the period-1 points of an arbitrary $z$-axis rotation, Fig.~\ref{fig:P1_schematic}(a) shows the image (solid red) of the initial tracer curve $x=0,\, y^2+z^2=1$ (vertical black) after the $z$-axis rotation. The points where the image intersects the original satisfy $M_{\theta_z}^z (\bm{x})= \bm{x}$, and hence are period-1 points of $M_{\theta_z}^z$, as are all the points on the same latitudes (dashed red). Similarly, the intersections of the curve $z=0,\, x^2+y^2=1$ [horizontal black in Fig.~\ref{fig:P1_schematic}(b)] with its image after an arbitrary $x$-axis rotation (solid blue) indicate curves of constant latitude that are period-1 points of $M_{\theta_x}^x$ (dashed blue). Points where the dashed curves in Fig.~\ref{fig:P1_schematic}(a) and Fig.~\ref{fig:P1_schematic}(b) intersect are period-1 points of both the $z$- and $x$-axis rotations, and hence are period-1 points for this particular BST protocol. Given an arbitrary protocol $(\theta_z,\theta_x,h)$, the position $(x,y,z)$ on the unit sphere with $y<-h$ is a period-1 point if and only if there exist integers $m,n$ such that 
\begin{align} %\label{eq:period-1_angles}
\theta_z &= 2m \arccos\left( h/\sqrt{1-z^2}  \right), \label{eq:period-1_angles_1} \\
\theta_x &= 2n \arccos\left( h/\sqrt{1-x^2}  \right), \label{eq:period-1_angles_2}
\end{align}
where $m,n$ represent the number of times a tracer particle seeded at $(x,y,z)$ crosses the flowing layer during the $z$- and $x$-axis rotations respectively. It follows that if $(x,y,z)$ is a period-1 point for the protocol $(\theta_z,\theta_x,h)$, then the points $(\pm x, y, \pm z)$ are also period-1 [see Fig.~\ref{fig:P1_schematic}(c)], and so there is a period-1 point in each quadrant of the hemisphere.

\begin{figure*}[tbp]
\centering
\includegraphics[width=\textwidth]{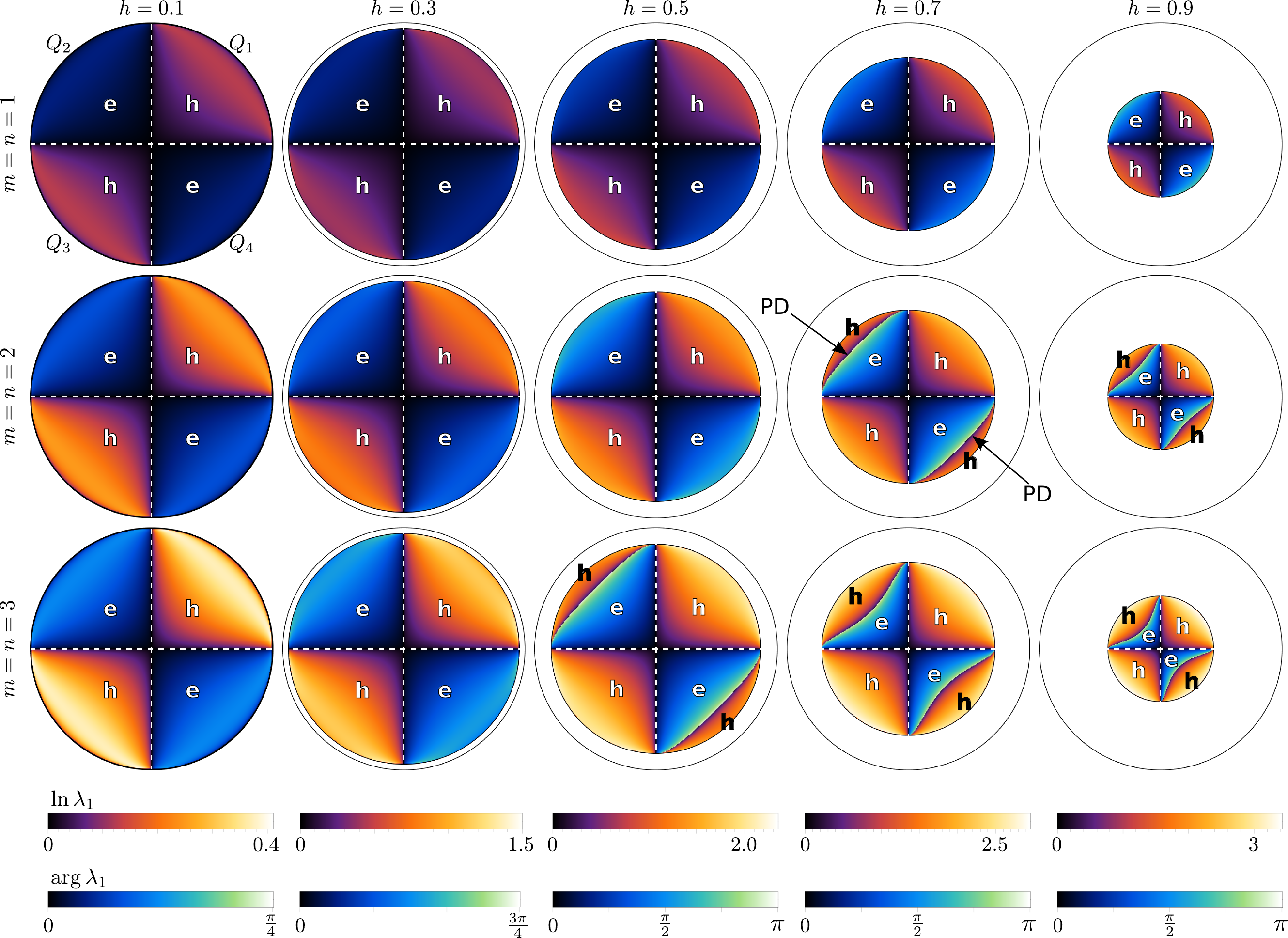}
\caption{Periodic point stability vs.\ position on the unit hemispherical shell (bottom view) for various $h$ (columns) and numbers of flowing layer crossings $m=n$ (rows) during each rotation ($z$- and $x$-axis). For each position $(x,y,z)$ on the unit hemispherical shell, eqs.~(\ref{eq:period-1_angles_1},\ref{eq:period-1_angles_2}) give the corresponding rotation angles $\theta_z,\theta_x$ such that it is a period-1 point. The eigenvalues $\lambda_{1,2,3}$ of the Jacobian $DM$ determine the local transport stability, assuming $\lambda_1 \geq 1$ if all the eigenvalues are real, and $\arg \lambda_1 > 0$ otherwise. If all eigenvalues are real, the period-1 point is hyperbolic, and the position $(x,z)$ is colored based on $\ln \lambda_1$ as indicated by the corresponding color bars (lighter means more expansion and contraction). If there is a complex conjugate pair of eigenvalues, the period-1 point is elliptic, and the position $(x,z)$ is colored blue/green based on $\arg \lambda_1$ as indicated by the corresponding color bars (lighter means a greater rotation angle). Note the varying color scales with increasing $h$. Period-1 points in $Q_1$ and $Q_3$ are always hyperbolic (marked as `h'). Period-1 points in $Q_2$ and $Q_4$ are elliptic (marked as `e'), unless period-doubling (PD) bifurcations occur (row 2, columns 4 and 5, and row 3 columns 3, 4 and 5), resulting in hyperbolic points near the edge of the domain. Period-1 points in the planes $x=0$ and $z=0$ (dashed white vertical and horizontal lines) are always degenerate (parabolic).}
\label{fig:hyperbolic_elliptic}
\end{figure*}

The stability of local particle transport near each period-1 point can be determined by analytically computing the Jacobian $D M = (\partial M_i / \partial x_j)$ and its eigenvalues $\lambda_{1,2,3}$. Since $M$ is a volume-preserving map, the product $\lambda_1 \lambda_2 \lambda_3 =1$, and since tracer particles are confined to 2D spherical shells \cite{Park2016}, one of the eigenvalues, say $\lambda_3$, must equal $1$ \cite{Anderson1999, Malyuga2002, Speetjens1}. If $\lambda_1 = 1/\lambda_2>1$ are real, then tracer particle transport in the neighborhood of the periodic point is unstable, and the periodic point is hyperbolic. These points are characterized by their exponent of expansion/contraction, $\ln \lambda_1>0$. Conversely, if $\lambda_{1,2} = \exp(\pm i\alpha)$, $\alpha>0$, then local particle transport is stable, and the periodic point is elliptic. These points are characterized by the rotation generated at the periodic point, $\alpha = \arg \lambda_1$. The final possibility is that $\lambda_1=\lambda_2=\pm 1$, in which case the periodic point is degenerate (or parabolic), and local transport is on the cusp of the stable/unstable paradigm. As a result of the combination of shears, the period-1 points in the quadrants $Q_1:x>0,z>0$ and $Q_3:x<0,z<0$ are hyperbolic for all rotation angles $\theta_z,\theta_x>0$ and all $0<h<1$. This is illustrated in Fig.~\ref{fig:hyperbolic_elliptic}, which shows the stability [red/purple: hyperbolic (`h'), blue/green: elliptic (`e')] of the period-1 point at each position on the unit hemispherical shell when $\theta_z,\theta_x$ satisfy eqs.~(\ref{eq:period-1_angles_1},\ref{eq:period-1_angles_2}), where the number of crossings $m=n$ corresponds to the row (e.g.,\ second row: $m=n=2$). It is clear that the period-1 points in quadrants $Q_1,Q_3$ are always hyperbolic (red/purple, `h'), with greater expansion and contraction factor, $\ln \lambda_1$, (lighter) nearer the domain boundary $\partial S$. At a given location $(x,z)$ in $Q_1$ or $Q_3$, $\ln \lambda_1$ increases as $h$ increases (going along each row in Fig.~\ref{fig:hyperbolic_elliptic}, noting the change in color scale for each column), and $\ln \lambda_1$ also increases as $m$ and $n$ increase (going down each column). Conversely, in the quadrants $Q_2:x<0,z>0$ and $Q_4:x>0,z<0$, the combination of shears is such that for small $h$ and low numbers of flowing layer crossings (toward the top left of Fig.~\ref{fig:hyperbolic_elliptic}), the period-1 points are mostly elliptic (blue/green, `e'), where lighter coloring indicates greater rotation generated at the elliptic point, $\arg \lambda_1$. In fact, it can be shown that if there is a single flowing layer crossing during each rotation phase ($m=n=1$, first row), then the period-1 points in $Q_2,Q_4$ are elliptic for all protocols $(\theta_z,\theta_x,h)$. However, if several flowing layer crossings are permitted ($m=n\geq 2$), and $h$ is sufficiently large, there exist protocols for which the period-1 points in $Q_2,Q_4$ are hyperbolic. Period-doubling (PD) bifurcations occur along the curves in $Q_2,Q_4$ where the stability changes from elliptic, with $\arg \lambda_1=\pi$, to hyperbolic with $\ln \lambda_1=0$ (e.g.\ second row, fourth column in Fig.~\ref{fig:hyperbolic_elliptic}). The proportion of elliptic points decreases as $h$ or $m=n$ increase. Note that period-1 points in the planes $x=0$ and $z=0$ (the dashed white vertical and horizontal lines) are always degenerate (or parabolic). While not shown in Fig.~\ref{fig:hyperbolic_elliptic}, the period-1 points for $m\neq n$ share the same properties, i.e. always hyperbolic in $Q_1,Q_3$, and generally elliptic in $Q_2,Q_4$, except when period-doubling bifurcations occur.

\begin{figure}[tbp]
\centering
\includegraphics[width=\columnwidth]{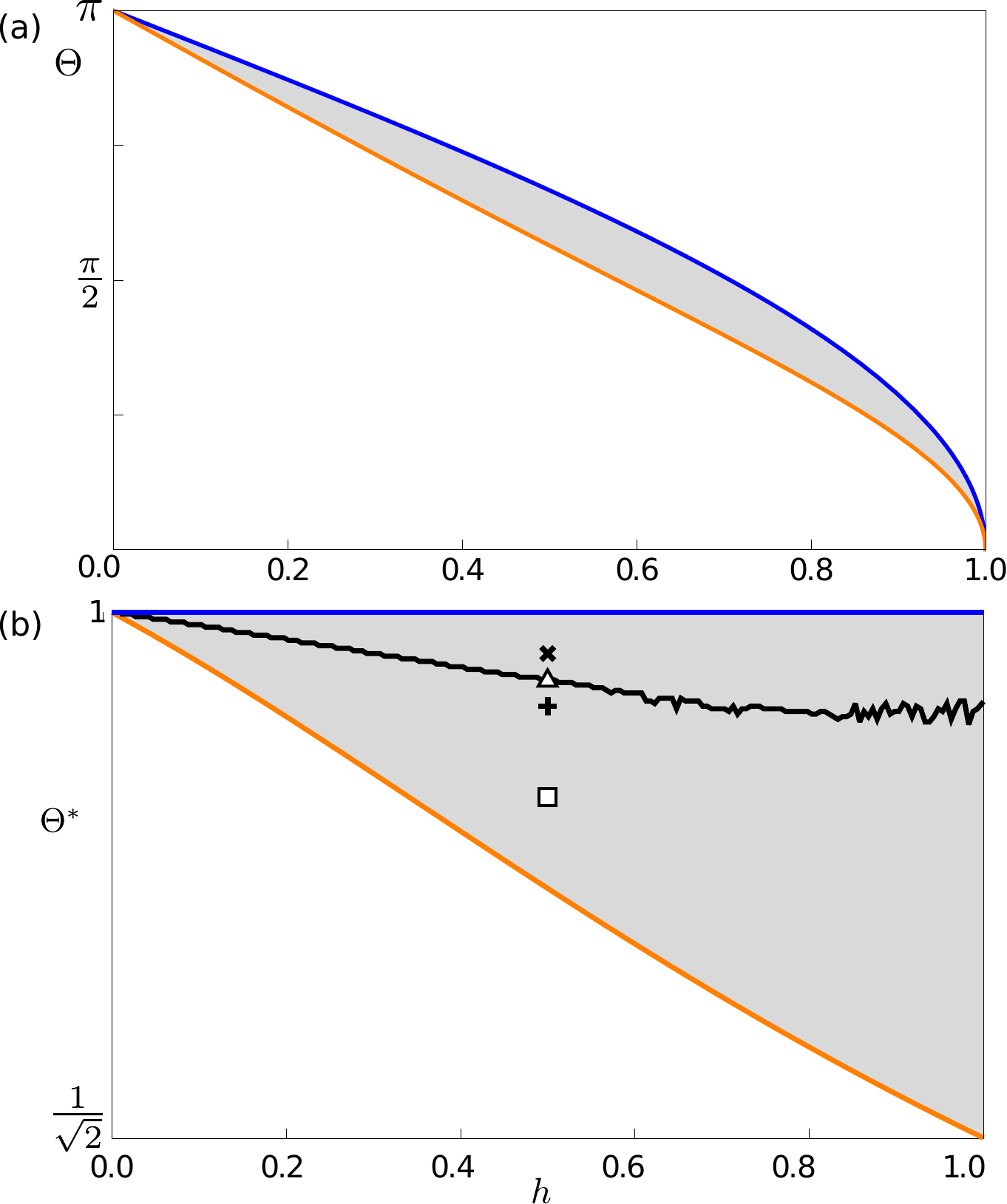}
\caption{(a)~For $\theta_z=\theta_x=\Theta$, protocols in the gray region exhibit period-1 points that cross the flowing layer once during each rotation ($z$- and $x$-axis). Along the blue (upper) curve, $\Theta = 2\arccos h$, there is a single period-1 point at $(0,-1,0)$ that is degenerate (parabolic). Along the orange (lower) curve, $\Theta= 2\arccos(h/\sqrt{(1+h^2)/2})$, there are four periodic points $(\pm \sqrt{(1-h^2)/2}, -h, \pm \sqrt{(1-h^2)/2})$ where the domain boundary $\partial S$ intersects the planes $z=\pm x$. (b)~The same as (a) except the rotation angle is normalized by the subtended angle between opposite points on $\partial S$ and the origin (i.e.\ the blue curve), $\Theta^* = \Theta/(2\arccos h)$. Along the black curve, $\Theta^*=\Theta^*_\text{crit}(h)$, inside the gray region, containment bifurcations occur, as detailed in \S\ref{sec:containment_bifurcation}.}
\label{fig:existence_region}
\end{figure}

In this paper we focus on the period-1 points that cross the flowing layer once ($m=n=1$) during each rotation phase, however, the same approach can be applied for multiple flowing layer crossings. If $h=0$, the map is a PWI, and from eqs.~(\ref{eq:period-1_angles_1},\ref{eq:period-1_angles_2}), a point $(x,y,z)$ is a period-1 point if and only if $\theta_z=\theta_x = 2\arccos(0)=2\pi$. Hence, either no points are period-1 or the entire domain is period-1. When the rotation angles are equal, $\theta_z=\theta_x=\Theta$, and $h\neq 0$, eqs.~(\ref{eq:period-1_angles_1},\ref{eq:period-1_angles_2}) are only satisfied when $z^2=x^2$, i.e.\ $z=\pm x$, and there are either four period-1 points at $(\pm x, \sqrt{1-2x^2}, \pm x)$ [e.g.\ Fig.~\ref{fig:P1_schematic}(d)], where 
\begin{equation}
x = \sqrt{1-\left(\frac{h}{\cos(\Theta/2)} \right)^2},
\end{equation}
or there is a single period-1 point at $(x,y,z)=(0,-1,0)$, when $\Theta= 2\arccos h$. For fixed $h \neq 0$, as $\Theta$ decreases, starting from $2\arccos h$ [moving vertically downward from the blue (upper) curve in Fig.~\ref{fig:existence_region}(a)], the period-1 point at $(0,-1,0)$ splits into four period-1 points, like those shown in Fig.~\ref{fig:P1_schematic}(d). As $\Theta$ continues to decrease, the four period-1 points move outward in the planes $z=\pm x$, until they eventually meet the boundary of the domain $\partial S$, where they annihilate. The boundary $\partial S: y=-h,\, x^2+y^2+z^2=1$ intersects the planes $z=\pm x$ when $2x^2 + h^2 =1$, i.e.\ at the points $(\pm \sqrt{(1 - h^2)/2},-h,\pm \sqrt{(1-h^2)/2})$. Therefore, from eqs.~(\ref{eq:period-1_angles_1},\ref{eq:period-1_angles_2}), the four period-1 points annihilate when $\Theta = 2\arccos(h/\sqrt{1-x^2}) = 2\arccos(h/\sqrt{(1+h^2)/2})$ [the orange (lower) curve in Fig.~\ref{fig:existence_region}(a)]. Thus, period-1 points only exist when $2\arccos(h/\sqrt{(1+h^2)/2}) \leq \Theta \leq 2\arccos h$, i.e.\ in the gray region between the blue and orange curves in Fig.~\ref{fig:existence_region}(a). As previously explained, for $h=0$ either no point in the domain is period-1 (when $\Theta \neq \pi$) or the entire domain is period-1 (when $\Theta = \pi$), meaning the blue and orange curves meet at $\Theta=\pi$ in the limit $h \to 0$ [the vertical axis in Fig.~\ref{fig:existence_region}(a)]. To non-dimensionalize the tumbler rotations, we normalize $\Theta$ by the subtended angle between any two diametrically opposed points on $\partial S$ and the origin, i.e.\ $\Theta^* =\Theta/(2\arccos h)$, which is also the maximum geodesic distance between any two points in the domain. The birth (upper blue) and annihilation (lower orange) curves of the period-1 points are shown in Fig.~\ref{fig:existence_region}(b) after normalization. The blue curve  becomes $\Theta^*=1$ for all $h$, and the orange curve monotonically decreases from $\Theta^*=1$ at $h=0$ to $\Theta^*=1/\sqrt{2}$ in the limit as $h\to 1$.

\section{Interplay between stretching-and-folding and cutting-and-shuffling actions} \label{sec:bifurcations}

\subsection{Elliptic island containment bifurcation} \label{sec:containment_bifurcation}

\begin{figure*}[tbp]
\centering
\includegraphics[width=0.8\textwidth]{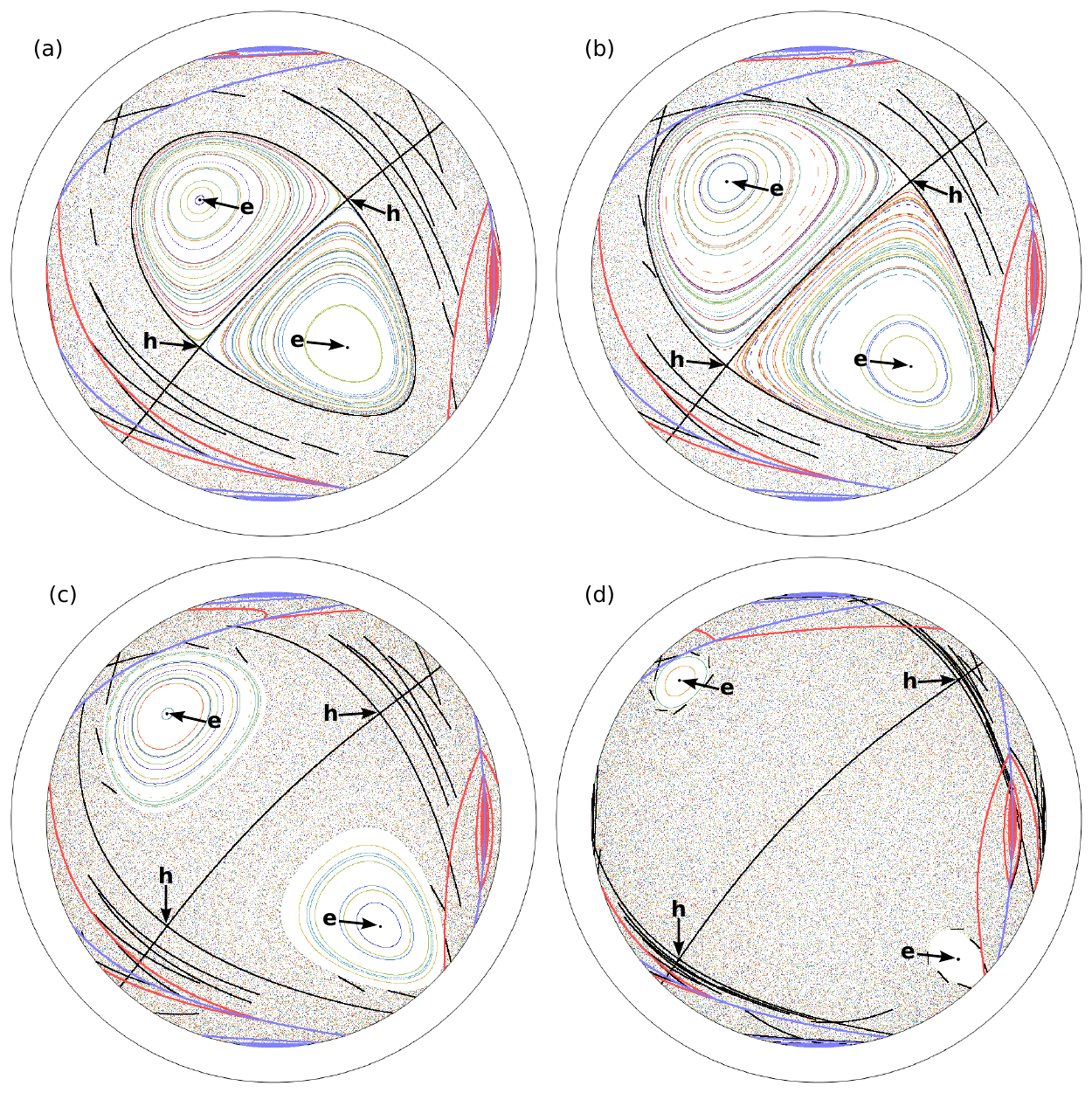}
\caption{Bifurcation from containment of the elliptic island in $Q_2$ by manifolds (thin black curves) to containment by cutting lines [thick blue and red (gray) curves], with equal rotation angles $\theta_z=\theta_x=\Theta=2 \Theta^* \arccos h$ and $h=0.5$. Period-1 points (elliptic: `e', or hyperbolic: `h') are marked with black points and arrows. (a)~For $\Theta^*=0.9764 >\Theta^*_\text{crit}$ [marked by the `$\times$' in Fig.~\ref{fig:existence_region}(b)], the elliptic islands are contained within the unstable and stable manifolds associated with the hyperbolic period-1 points (same as Fig.~\ref{fig:containment_psection_1}). (b)~For $\Theta^* =0.9621 =\Theta^*_\text{crit}$ [triangle in Fig.~\ref{fig:existence_region}(b)], the manifolds enclosing the island in $Q_2$ are tangent to a cutting line. (c)~For $\Theta^* =0.9473 < \Theta^*_\text{crit}$ [`+' in Fig.~\ref{fig:existence_region}(b)], the manifolds that enclosed the elliptic island in $Q_2$ are cut by a cutting line, and the elliptic island is limited in size by its distance to the nearest cutting line. (d)~$\Theta^* = 0.8967 \ll \Theta^*_\text{crit}$ [square in Fig.~\ref{fig:existence_region}(b)], as $\Theta^*$ decreases the period-1 points move outward and the elliptic islands shrinks.}
\label{fig:containment_bifurcation}
\end{figure*}

Returning to the example briefly discussed in \S\ref{sec:intro}, here we describe the bifurcation from Lagrangian transport structures controlled by stretching-and-folding actions to structures controlled by cutting-and-shuffling actions. To demonstrate, consider the case $h=0.5$ and $\Theta^*=0.9764$ which was shown in Fig.~\ref{fig:containment_psection_1}, and is again shown in Fig.~\ref{fig:containment_bifurcation}(a), corresponding to the point marked by the black `$\times$' in Fig.~\ref{fig:existence_region}(b). In this case the elliptic islands are confined by the stable and unstable manifolds (thin black curves) associated with the period-1 hyperbolic points (marked by `h'), as occurs in classical stretching-and-folding systems. As $\Theta^*$ decreases, the periodic points move outward, and the elliptic islands grow. At a critical value, $\Theta^*_{\text{crit}}=0.9621$ [Fig.~\ref{fig:containment_bifurcation}(b) corresponding to the triangle in Fig.~\ref{fig:existence_region}(b)], the perimeter of the island in $Q_2$, and the manifolds that contain it, become tangent to a thick blue (gray) cutting line. For $\Theta^*$ less than $\Theta^*_\text{crit}$, the size of the elliptic island is limited by the distance to the nearest cutting line, as occurs in cutting-and-shuffling systems. Furthermore, as $\Theta^*$ decreases below $\Theta^*_\text{crit}$, the periodic points move outward. Hence, the island shrinks, as shown for $\Theta^*=0.9473, 0.8967$ in Fig.~\ref{fig:containment_bifurcation}(c,d), corresponding to the `+' and square in Fig.~\ref{fig:existence_region}(b) respectively. Therefore, the critical value $\Theta^*_\text{crit}$ represents a bifurcation from containment of the elliptic island in $Q_2$ by manifolds (stretching-and-folding structures) to containment by cutting lines (cutting-and-shuffling structures). Note that the same bifurcation occurs for both the islands, in $Q_2$ and $Q_4$, though the bifurcation points are different. The bifurcation occurs when the manifolds, which form barriers to transport above the critical value, $\Theta^*_\text{crit}$, are cut, and tracer particles are able to pass between the cut segments. Therefore, particles in the chaotic region are able to enter the region between the remaining elliptic island and the edge of the manifolds.

\begin{figure}[tbp]
\centering
\includegraphics[width=\columnwidth]{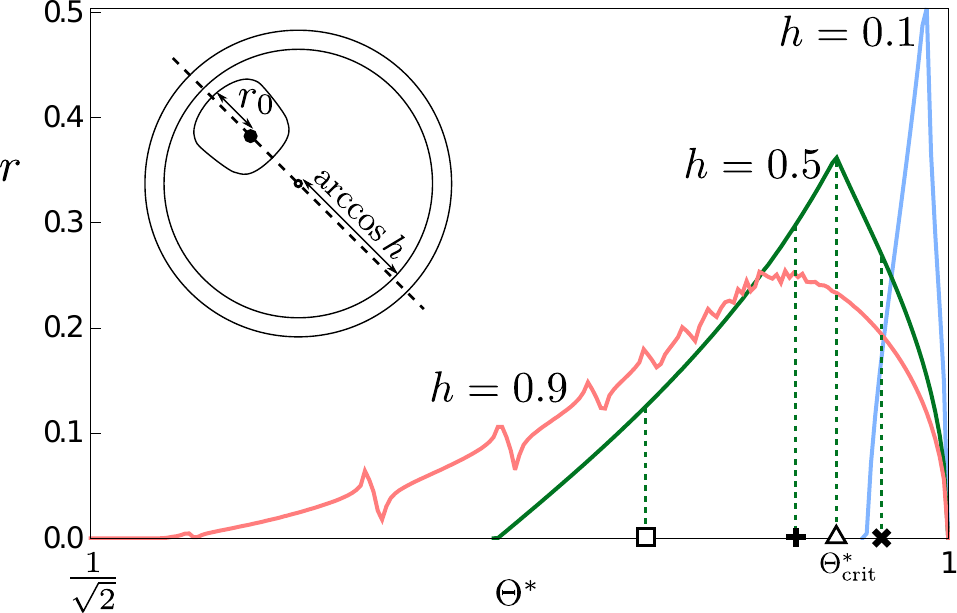}
\caption{Normalized radius, $r=r_0/\arccos h$ of the elliptic region in quadrant $Q_2$, vs. normalized rotation angle $\Theta^*$ at different values of $h$. Shown inset, $r_0$ is the geodesic distance from the period-1 point to the perimeter of the island along the curve $\mathcal{C}: z=-x,\, x^2+y^2+z^2=1$ (dashed) in the negative $x$ direction. The sharp maxima in $r$ correspond to bifurcations from containment by manifolds to containment by cutting lines. For $h=0.5$ (dark-green curve), values of $\Theta^*$ corresponding to Fig.~\ref{fig:containment_bifurcation}(a--d) are indicated by the `$\times$', triangle, `+', and square, respectively, as in Fig.~\ref{fig:existence_region}(b).}
\label{fig:radius_sampled}
\end{figure}

To describe this containment bifurcation in more detail, we define the normalized radius $r$ of the elliptic island in $Q_2$ as the distance $r_0$ from the period-1 elliptic point to the boundary of the elliptic island along the symmetry curve $\mathcal{C}:z=-x,\, x^2+y^2+z^2=1$ \footnote{For equal rotation angles, $\theta_z=\theta_x$, the BST map possesses a reflection-reversal symmetry \cite{Smith2017BSTresonances}, meaning invariant structures, e.g.\ islands and manifolds, must be symmetric about the plane $z=-x$.} in the negative $x$ direction [see inset in Fig.~\ref{fig:radius_sampled}] scaled by half the maximal geodesic distance between any two points in the domain, $r=r_0/\arccos h$. While not exactly the same, the same approach can be applied to define the radius of the elliptic island in $Q_4$. We use $r$ to quantify the size of the island, and relate it to the protocol $(\Theta^*,h)$. Fig.~\ref{fig:radius_sampled} shows that for each fixed value of $h$, the radius $r$ exhibits a sharp transition from increasing to decreasing at its maximum value. 

Consider, for example, the dark-green curve in Fig.~\ref{fig:radius_sampled}, corresponding to $h=0.5$ for which an example of the bifurcation is shown in Fig.~\ref{fig:containment_bifurcation}. The maximum of the dark-green curve corresponds to $\Theta^*_\text{crit}$ in Fig.~\ref{fig:containment_bifurcation}(b) for which the cutting lines are tangent to the manifolds containing the elliptic islands. To the right of $\Theta^*_\text{crit}$, the cutting lines lie outside of the manifolds containing the elliptic island, as in Fig.~\ref{fig:containment_bifurcation}(a). The point where the dark-green curve for $h=0.5$ in Fig.~\ref{fig:radius_sampled} reaches $\Theta^*=1$ corresponds to the point on the blue (upper) curve in Fig.~\ref{fig:existence_region}(b) at $h=0.5$, which is where a single period-1 point first appears at $(x,y,z)=(0,-1,0)$. To the left of the maximum of the dark-green curve in Fig.~\ref{fig:radius_sampled}, the cutting lines reduce the size of the elliptic islands as the elliptic points move outward, examples of which are shown in Fig.~\ref{fig:containment_bifurcation}(c,d). As $\Theta^*$ decreases from $\Theta^*_\text{crit}$, the elliptic points eventually move outward far enough that they reach the boundary $\partial S$.  This occurs where the dark-green curve for $h=0.5$ in Fig.~\ref{fig:radius_sampled} reaches $r=0$ to the left of $\Theta^*_\text{crit}$, which corresponds to the value for $\Theta^*$ on the orange (lower) curve in Fig.~\ref{fig:existence_region}(b) for $h=0.5$. Thus, the curves in Fig.~\ref{fig:radius_sampled} reflect the size of the islands due to the combination of cutting-and-shuffling and stretching-and-folding along a vertical line in Fig.~\ref{fig:existence_region}(b). Elliptic islands for values of $\Theta^*$ above the black curve in the gray region of Fig.~\ref{fig:existence_region}(b) are unaffected by the cutting lines, while elliptic islands for values of $\Theta^*$ below the black curve are reduced in size due to the cutting lines, eventually being annihilated completely when the corresponding periodic points reach $\partial S$ (lower orange curve).

\begin{figure}[tbp]
\centering
\includegraphics[width=\columnwidth]{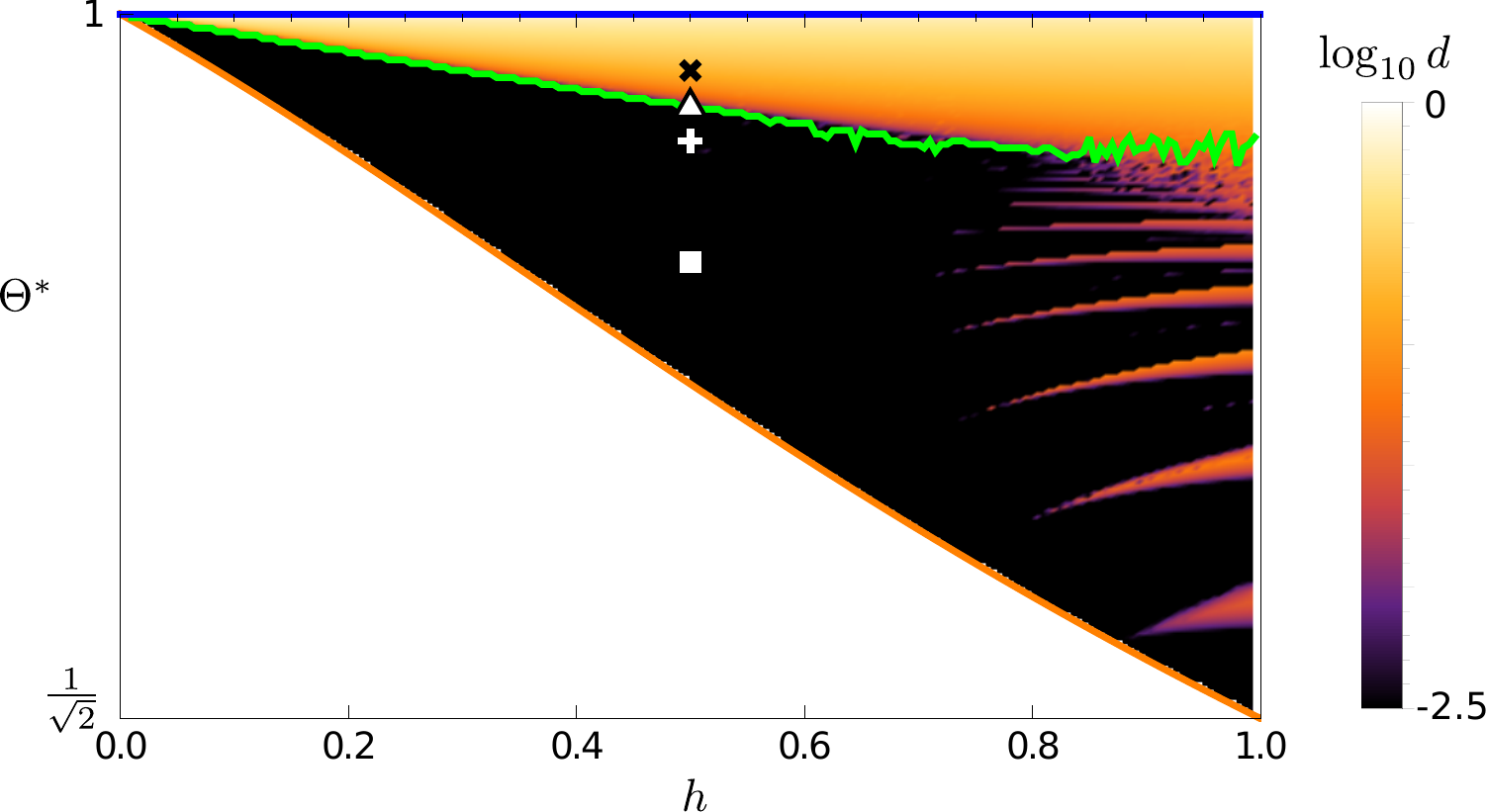}
\caption{Normalized distance, $d$, from the perimeter of the elliptic region in quadrant $Q_2$ to the nearest cutting line (normalized by half the maximal geodesic distance between points in the domain, $\arccos h$). Tumbler rotation angles $\Theta^*_\text{max}$ corresponding to the maximal radius $r$ for each value of $h$ are shown by the light green curve which separates the light and dark regions. Since $d$ is computed numerically, it is never exactly zero. However, in the black region $d$ is close to zero, meaning the island is tangent, or almost tangent, to a cutting line. Conversely, protocols with non-zero $d$ (light-colored) correspond to islands that are contained by manifolds. Since the curve $\Theta^*_\text{max}$ separates the regions $d=0$ (black) and $d\neq 0$ (light-colored), it also corresponds to the bifurcation curve, i.e. $\Theta^*_\text{max}=\Theta^*_\text{crit}$. Protocols $(\Theta^*,h)$ corresponding to the Poincar\'{e} sections Fig.~\ref{fig:containment_bifurcation}(a--c) are indicated by the black `$\times$', black-outlined white triangle, white `+', and white square, respectively, as in Fig.~\ref{fig:existence_region}(b).}
\label{fig:radius+distance2cut}
\end{figure}

To show that the sharp maxima in Fig.~\ref{fig:radius_sampled} correspond to the containment bifurcation points more generally (not just for $h=0.5$), we consider the normalized distance from the perimeter of the elliptic island to the nearest cutting line, $d$ \footnote{Like the radius $r$, $d$ is normalized by scaling by half the maximal geodesic distance between any points in the domain, $\arccos h$.}, whose distribution across the $(h,\Theta^*)$ parameter space is shown in Fig.~\ref{fig:radius+distance2cut}. It is clear that the curve $\Theta^*_{\text{max}}(h)$ corresponding to the maximum radius for each value of $h$ (light green) divides the parameter space into two regions with vastly different phenomena. For $\Theta^*>\Theta^*_{\text{max}}$, the distance $d$ is non-zero (light-colored), meaning the elliptic island is contained by manifolds, as in Fig.~\ref{fig:containment_bifurcation}(a). Conversely, for $\Theta^*<\Theta^*_{\text{max}}$, aside from the colored bands evident for $h>0.7$, the distance $d \approx 0$ (black), meaning the island is tangent to a cutting line, as in Fig.~\ref{fig:containment_bifurcation}(c). Therefore, the rotation angle corresponding to maximal island radius also corresponds to the bifurcation from containment by manifolds to containment by cutting lines, i.e.\ $\Theta^*_\text{max}=\Theta^*_\text{crit}$. 

The colored bands observed in Fig.~\ref{fig:radius+distance2cut} for $h>0.7$ and the jaggedness of the curve $h=0.9$ in Fig.~\ref{fig:radius_sampled} are caused by another type of interaction between the elliptic island and the cutting lines, when chains of elliptic and hyperbolic periodic points originating within the period-1 elliptic island are annihilated. This will be discussed in detail in \S\ref{sec:annihilation_bifurcation}.

Recall that in all the cases described here, the flowing layer is infinitely thin. Considering instead the BST with a finite thickness flowing layer, or more generally, considering cutting-and-shuffling as a localized shear, the containment bifurcation will have an even greater impact, as the cutting lines become ``thickened''. Elliptic islands contained by manifolds are unaffected by the thickness of the flowing layer as long as the smooth deformations that govern them remain the same. However, elliptic islands tangent to a cutting line will shrink as the flowing layer thickness (equivalent to the cutting line width) increases. Therefore, in experimental studies of the BST flow and other systems that can be approximated with combined stretching-and-folding and cutting-and-shuffling actions, we expect elliptic islands contained by manifolds to be more robust than those contained by cutting lines, and hence provide a greater hindrance to chaotic particle transport. However, this issue is left for future investigations.

\subsection{Periodic point annihilation} \label{sec:annihilation_bifurcation}

In classical stretching-and-folding systems periodic points cannot annihilate unless they reach a domain boundary. However, the same is not true when cutting-and-shuffling motions are present. For instance, in PWIs elliptic periodic points can spontaneously appear and disappear. In essence, the cutting lines can be thought of as domain sub-boundaries, because they separate regions of distinct flow, and periodic points annihilate when they reach a cutting line. Below, we demonstrate that spontaneous annihilation of periodic points can also occur in the interior of the domain for the less than half full BST map.

\begin{figure*}[tbp]
\centering
\includegraphics[width=\textwidth]{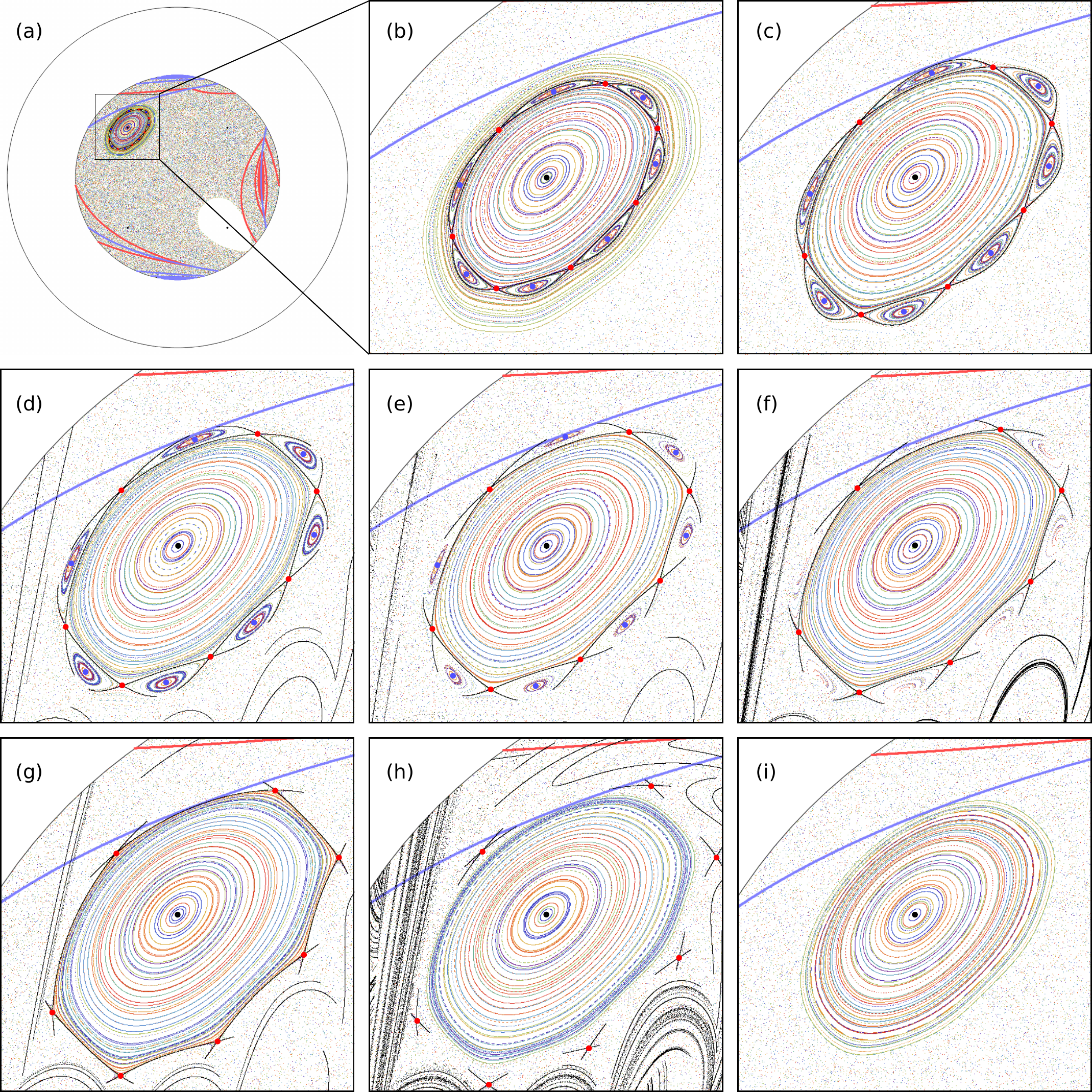}
\caption{As $\theta_z=\theta_x=\Theta=2\Theta^* \arccos h$ decrease, period-7 chains of elliptic and hyperbolic periodic points (predicted by the Poincar\'{e}--Birkhoff theorem) annihilate when they meet the thick blue (gray) cutting line (bottom view). The distance $h=0.8$ is kept constant. (a)~Bottom view of a Poincar\'{e} section for $\Theta^*= 0.89982$. Cutting lines are shown as thick blue and red (gray) curves, period-1 points are marked by black points. (b)~A zoomed in box around the period-1 point in quadrant $Q_2$. Alternating chains of elliptic (blue) and hyperbolic (red) period-7 points exist within the perimeter of the island. Unstable and stable manifolds associated with the period-7 hyperbolic points are shown black, forming parallel heteroclinic connections. (c)~At $\Theta^*=0.89735 =\Theta^*_1$ the outer manifolds, defining the island perimeter, are tangent to a cutting line. (d)~For $\Theta^* =0.89712  < \Theta^*_1$, the outer manifold is cut, allowing the chaotic region to enter between the period-1 island and the period-7 islands. The inner manifold is now the perimeter of the period-1 island, and is not tangent to a cutting line. (e)~$\Theta^*=0.89673$, similar to (d). (f)~At $\Theta^*=0.89634 =\Theta^*_2$, one of the period-7 elliptic points meets a cutting line, and all the period-7 elliptic points annihilate. (g)~At $\Theta^*=0.89469 =\Theta^*_3$, the inner manifolds are tangent to a cutting line, meaning the distance $d$ from the cutting lines to the island is again zero. (h)~For $\Theta^*=0.89358 <\Theta^*_3$, the inner manifolds are also cut, and no longer form the perimeter of the island. (i)~At $\Theta^*=0.89117 =\Theta^*_4$, one of the period-7 hyperbolic points meets a cutting line, and all the period-7 hyperbolic points annihilate, leaving only the period-1 island.}
\label{fig:PB_island_destruction}
\end{figure*}

\begin{figure}[tbp]
\centering
\includegraphics[width=\columnwidth]{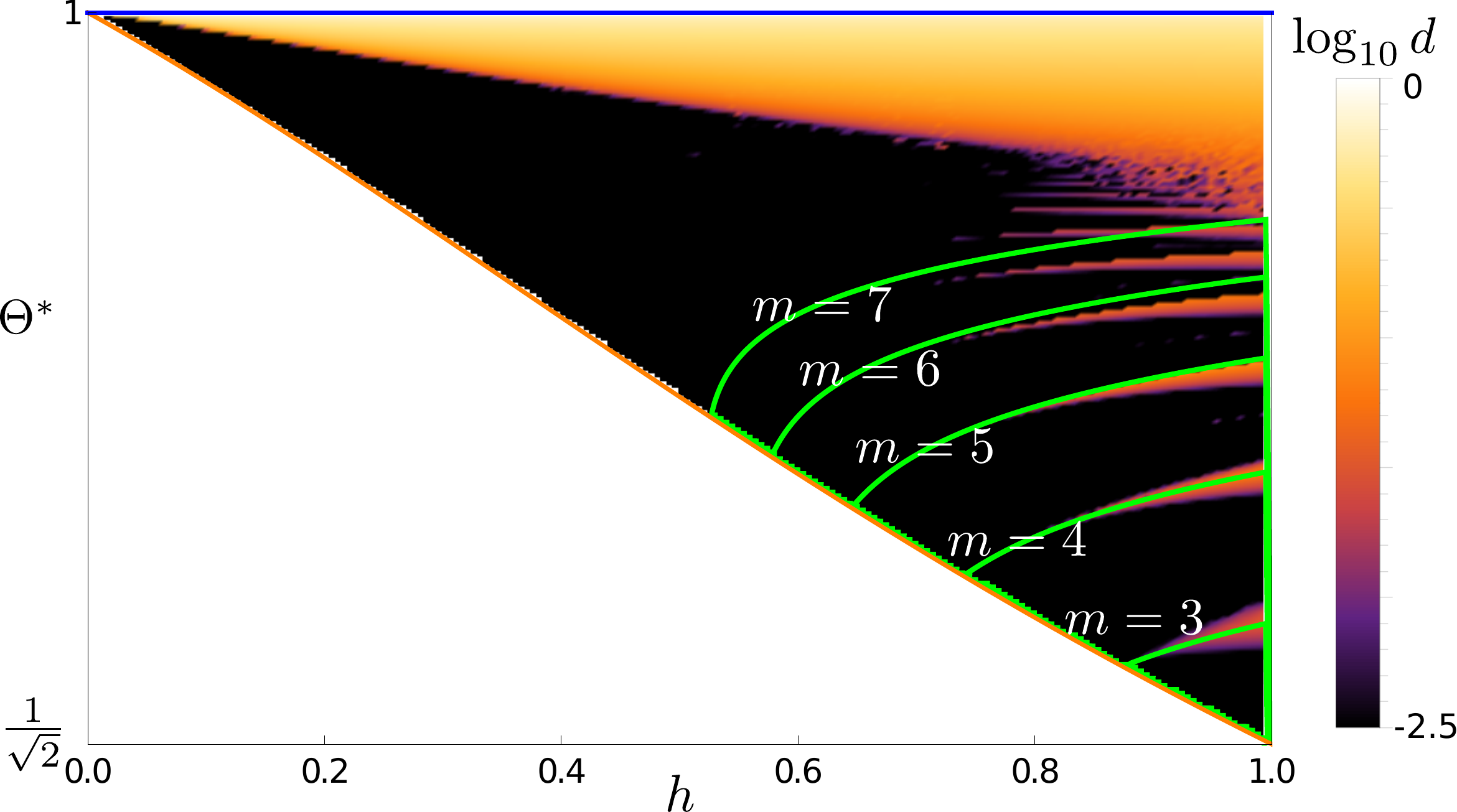}
\caption{The distance $d$ from the edge of the elliptic region to the nearest cutting line (Fig.~\ref{fig:radius+distance2cut}) overlayed with light green contours $\alpha = 2\pi/m$, $m=3,4,\dots,7$, where $\alpha$ is the rotation angle at the period-1 elliptic point ($\lambda_{1,2} = \exp (\pm i \alpha)$). Each light green curve indicates the presence of alternating chains of elliptic and hyperbolic period-$m$ points (predicted by the Poincar\'{e}--Birkhoff Theorem) for nearby protocols. These chains of elliptic and hyperbolic points give rise to the light-colored bands corresponding to $d>0$ and observed for $h>0.7$, through the same manifold cutting process demonstrated in Fig.~\ref{fig:PB_island_destruction}(c,d).}
\label{fig:rotation_angle_overlaying_d2cut}
\end{figure}

By definition, the region within the perimeter of an elliptic island can never be cut. Therefore, tracer particles within an elliptic island only experience smooth deformation, and the classical theory of smooth area-preserving maps applies. In particular, KAM-theory and the Poincar\'{e}--Birkhoff theorem predict that invariant tori with low-order rational winding numbers surrounding an elliptic periodic point will break up into chains of alternating elliptic and hyperbolic periodic points. This is demonstrated by the period-7 elliptic and hyperbolic points in Fig.~\ref{fig:PB_island_destruction}(a--c), where the heteroclinic connections of the unstable and stable period-7 manifolds (black) contain the period-7 elliptic regions. As $\Theta^*$ decreases, the period-7 points move radially outward, away from the central period-1 point. At the same time, the period-1 point moves slightly closer to the thick blue (gray) cutting line. At a critical value, $\Theta^*_1$, the outer period-7 manifolds become tangent to the thick blue (gray) cutting line [Fig.~\ref{fig:PB_island_destruction}(c)]. At this critical value, the outer manifolds form the perimeter of the island. For $\Theta^*< \Theta^*_1$, the outer manifolds are cut where they intersect the cutting lines, and tracer particles from the chaotic region are able to leak into the regions surrounding the elliptic period-7 points [Fig.~\ref{fig:PB_island_destruction}(d,e)]. At these values, $\Theta^*<\Theta^*_1$, the inner heteroclinic manifold connections form the perimeter of the period-1 island, meaning the period-1 island has detached from the cutting line, and the distance $d$ from the island to the cutting line is positive. As $\Theta^*$ continues to decrease, the distance between the elliptic period-7 points and the cutting line also decreases. At a second critical value, $\Theta^*_2$, one of the period-7 elliptic points meets the cutting line [Fig.~\ref{fig:PB_island_destruction}(f)], and all the period-7 elliptic points annihilate. The inner heteroclinic manifold connections remain intact, forming the perimeter of the period-1 island until a third critical value $\Theta^*_3$ [Fig.~\ref{fig:PB_island_destruction}(g)] at which one of the inner manifolds becomes tangent to the cutting line, and the period-1 island reattaches to the cutting line. For $\Theta^*< \Theta^*_3$, both the inner and outer period-7 manifolds are cut by the cutting line [Fig.~\ref{fig:PB_island_destruction}(h)], and neither form the perimeter of the period-1 island (which remains attached to the cutting line). Decreasing $\Theta^*$ further, the period-7 hyperbolic points also move radially outward away from the period-1 point and toward the cutting line, and eventually annihilate when they reach the cutting line [Fig.~\ref{fig:PB_island_destruction}(i)]. 

This annihilation of chains of elliptic and hyperbolic points is responsible for the jagged curve in Fig.~\ref{fig:radius_sampled} for $h=0.9$ in that it generates the bands evident for $0.7< h <1, \, \Theta^* < \Theta^*_\text{max}$ in Fig.~\ref{fig:radius+distance2cut}. In Fig.~\ref{fig:PB_island_destruction}(b,c,g--i) the elliptic island is tangent to the cutting line, so $d=0$. However, for $\Theta^*_1< \Theta^* < \Theta^*_3$ [Fig.~\ref{fig:PB_island_destruction}(d--f)], the outer manifolds are cut, and the inner manifold, which is not tangent to the cutting line, forms the perimeter of the period-1 island, meaning $d>0$. This period-7 annihilation occurs for a range of $h$ values, resulting in one of the thin light-colored bands where $d>0$ below the light green curve separating light and dark regions in Fig.~\ref{fig:radius+distance2cut}. For any particular $h$, each of these bands results in a sharp change of slope in the curves in Fig.~\ref{fig:radius_sampled}, such as those appearing in the curve for $h=0.9$. 

Furthermore, chains of elliptic and hyperbolic points occur for many low-order periodicities. Referring back to the eigenvalues $\lambda_{1,2,3}$ of the Jacobian at the elliptic period-1 point, of the form $\lambda_{1,2} = \exp(\pm i \alpha)$, $\lambda_3=1$ for an elliptic periodic point. For protocols $(\Theta^*,h)$ with $\alpha = 2\pi/m$ for $m=3,4,5\dots$ (light green curves in Fig.~\ref{fig:rotation_angle_overlaying_d2cut}), the winding number at the periodic point is rational, and, for smaller values of $\Theta^*$, one of the invariant tori surrounding the elliptic point has the same rational winding number. According to KAM-theory and the Poincar\'{e}--Birkhoff theorem, tori with a low-order rational winding number generically break-up into alternating chains of elliptic and hyperbolic points, like that in Fig.~\ref{fig:PB_island_destruction}. Hence, for protocols near a protocol with $\alpha = 2\pi/m$, $m=3,4,5,\dots$, we observe similar bifurcation sequences to that seen in Fig.~\ref{fig:PB_island_destruction}, and observe a thin light-colored band where $d>0$ [Fig.~\ref{fig:rotation_angle_overlaying_d2cut}]. Note that the light green curves of constant rational winding number do not exactly overlay the thin bands, as they represent the rotation $\alpha$ at the center of the elliptic island, not at the perimeter. The tumbler rotation angle $\Theta^*$ must decrease sufficiently before the invariant torus at the perimeter of the island is rational, and the bifurcation sequence begins.

This annihilation of periodic points is expected to be a generic feature of systems with combined stretching-and-folding and cutting-and-shuffling actions. In particular, this occurs when an elliptic island is tangent to a cutting line and the chains of elliptic and hyperbolic periodic points arising from the Poincar\'{e}--Birkhoff theorem move away from the island center and collide with the cutting line.

\section{Conclusions} \label{sec:conclusions}

\begin{figure*}[tbp]
\centering
\includegraphics[width=\textwidth]{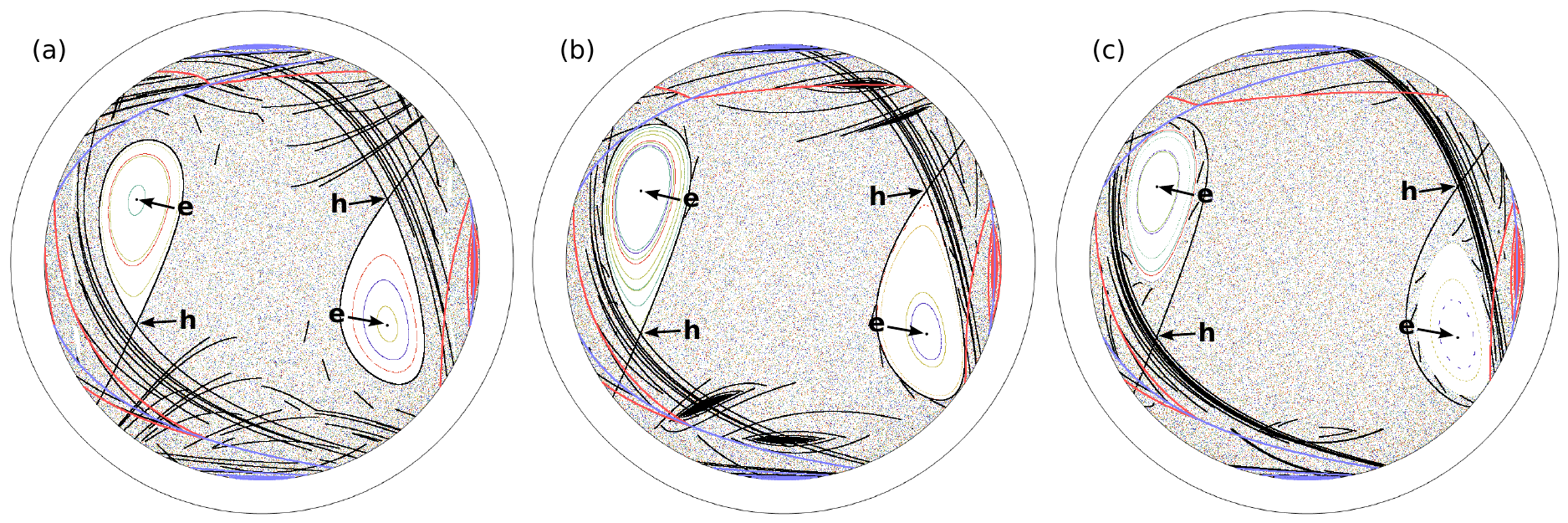}
\caption{Similar to Fig.~\ref{fig:containment_bifurcation}, bifurcation from containment of the elliptic islands by manifolds (thin black curves) to containment by cutting lines [thick blue and red (gray) curves] for unequal rotation angles in the less than half full BST with $h=0.5$. Period-1 points (elliptic: `e', or hyperbolic `h') are marked with black points. The angles $\theta_z,\theta_x$ are such that the four period-1 points are located in the planes $z=\pm x/2$, satisfying eqs.~(\ref{eq:period-1_angles_1},\ref{eq:period-1_angles_2}). (a)~For tumbler rotation angles $(\theta_z,\theta_x)=(2.056,1.911)$ [period-1 points located at $(x,z)=(\pm 0.5,\pm 0.25)$], the elliptic islands are contained within the unstable and stable manifolds associated with the hyperbolic period-1 points. (b)~For $(\theta_z,\theta_x)=(2.044,1.833)$ [period-1 points located at $(x,z)=(\pm 0.570,\pm 0.285)$], the manifolds enclosing the islands are tangent to a cutting line. (c)~For $(\theta_z,\theta_x)=(2.038,1.791)$ [period-1 points located at $(x,z)=(\pm 0.6,\pm 0.3)$], the manifolds that enclosed the elliptic islands are cut by the cutting lines, and the elliptic island is limited in size by its distance to the nearest cutting line. The chaotic region penetrates the regions between the islands and the manifolds.}
\label{fig:containment_bifurcation_non-eq}
\end{figure*}

The presence of concurrent stretching-and-folding and cutting-and-shuffling actions leads to novel transport structures and greater topological freedom for tracer particles. Classical dynamical systems theory for smooth maps and flows can be applied to structures that are wholly contained within a region that is not cut, but is not valid across the entire domain. In particular, the motion of tracer particles within non-mixing islands associated with elliptic periodic points is still governed by classical dynamical systems theory, including KAM-theory. This is observed in our model system, the BST map, where invariant tori whose rotation number is close to a low order rational number break up into alternating chains of elliptic and hyperbolic periodic points, as per the Poincar\'{e}--Birkhoff theorem. Of particular interest is when these classical structures meet and interact with the cuts. We have shown a generic case where chains of elliptic and hyperbolic points move radially away from the center of an elliptic island as a system parameter decreases. Since the perimeter of the elliptic island is tangent to a cutting line before the bifurcation sequence, when the chains of elliptic and hyperbolic points reach the perimeter of the island they begin to interact with the cutting line. At first, the outer manifolds are cut, removing a barrier to transport and allowing the chaotic region to leak between the central island and the chain of higher period elliptic islands. Eventually the elliptic periodic and hyperbolic periodic points are annihilated by the cut, meaning that the Poincar\'{e} index (a topological invariant of smooth systems) is not preserved.

We have also demonstrated a bifurcation between transport controlled by stretching-and-folding motions to transport controlled by cutting-and-shuffling motions. By decreasing the tumbler rotation about each axis (decreasing $\Theta^*$), the period-1 elliptic islands transition from containment by manifolds associated with period-1 hyperbolic points (a phenomenon associated with smooth systems) to containment by cutting line tangency (as occurs in PWIs). When the manifolds initially containing the island are cut, they no longer act as transport barriers, and tracer particles are able to enter the region between the manifolds and the island. The point of bifurcation, when the containing manifolds, island, and cutting lines are all tangent, represents a sharp maximum of the radius of the island in the BST map, indicating poor mixing performance. When the manifolds form transverse intersections rather than parallel connections, the manifolds still form barriers to transport that can contain islands. In these cases the same containment bifurcation can occur, where the containing manifolds are cut, and the size of the island becomes limited by its distance to a cutting line.

While the examples presented here are for equal tumbler rotation angles ($\theta_z=\theta_x$), the bifurcations between stretching-and-folding and cutting-and-shuffling controlled transport also occur for unequal rotation angles. For example, Fig.~\ref{fig:containment_bifurcation_non-eq} shows a similar bifurcation from confinement of the elliptic period-1 islands by manifolds [thin black curves in Fig.~\ref{fig:containment_bifurcation_non-eq}(a)] to confinement by cutting lines [thick red and blue (gray) curves in Fig.~\ref{fig:containment_bifurcation_non-eq}(c)] when the angles are not equal. In this case the angles $\theta_z,\theta_x$ are chosen such that the four period-1 points lie in the planes $z=\pm x/2$ using eqs.~(\ref{eq:period-1_angles_1},\ref{eq:period-1_angles_2}). Like in the equal angle case, the bifurcation occurs when the manifolds, island perimeter, and cutting lines all become tangent, as occurs in Fig.~\ref{fig:containment_bifurcation_non-eq}(b). Furthermore, the curve $\Theta^*_\text{crit}(h)$ that represents the containment bifurcation in the 2D parameter space $(\Theta^*,h)$ in Fig.~\ref{fig:radius+distance2cut} can be extended to a bifurcation surface in the full 3D parameter space $(\theta_z,\theta_x,h)$. More generally, the qualitative descriptions of the containment and annihilation bifurcations will be generic to the islands associated with all elliptic periodic points in systems with combined stretching-and-folding and cutting-and-shuffling.

Future work should focus on the evolution of these bifurcations and interactions away from the infinitely thin flowing layer limit. Christov \emph{et al.} \cite{Christov2014} showed that shear generated by a finite thickness flowing layer in the half-full case results in elliptic and hyperbolic period-1 structures similar to those observed here for the case when the flowing later is infinitely thin and the tumbler is less than half-full. Therefore, we expect that for a finite thickness flowing layer and a less than half-full tumbler, the period-1 structures should also be similar, meaning experiments such as those already performed for half-full tumblers \cite{Zaman2017} should also exhibit similar structures when less than half full. 

Furthermore, in the presence of a finite thickness flowing layer, the cutting-and-shuffling action is replaced with a localized shear, and the cutting lines are replaced by thickened regions representing the edges of the flowing layer \cite{Zaman2017}. Due to the high shear in the flowing layer, it is expected that tangent intersections between islands and the cutting lines in the infinitely thin limit will translate to tangent intersections between islands and the thickened cutting line. Consequently, islands tangent to cutting lines in the infinitely thin limit shrink as the flowing layer, and hence the cutting lines, are thickened. Alternatively, islands contained by manifolds should be comparatively more robust away from the infinitely thin limit, as they are unaffected by flowing layer thickening.

Other future directions include the consideration of non-orthogonal rotation axes and more than two rotation axes. In particular, having more than two rotation axes can break flow symmetries, creating the potential for better mixing. Another future direction is to consider other 3D tumbler geometries (e.g., ellipsoid, cuboid) which would result in 3D transport (not confined to 2D surfaces) via streamline jumping \cite{Christov2010, Christov2011streamline, Meier2007} in conjunction with both stretching-and-folding and cutting-and-shuffling. Understanding mixing and transport in these systems would provide further insight into industrial mixer designs. A deeper understanding of the interplay between stretching-and-folding and cutting-and-shuffling should open a wealth of possibilities for the design of novel mixing devices.

\begin{acknowledgments}
P.B. Umbanhowar was partially supported by the National Science Foundation Contract No. CMMI-1435065. 
\end{acknowledgments}

\appendix

\section{Equivalence between fill depth and spherical shell radius} \label{sec:LR_equiv}

\begin{figure}[b]
\centering
\includegraphics[width=0.7\columnwidth]{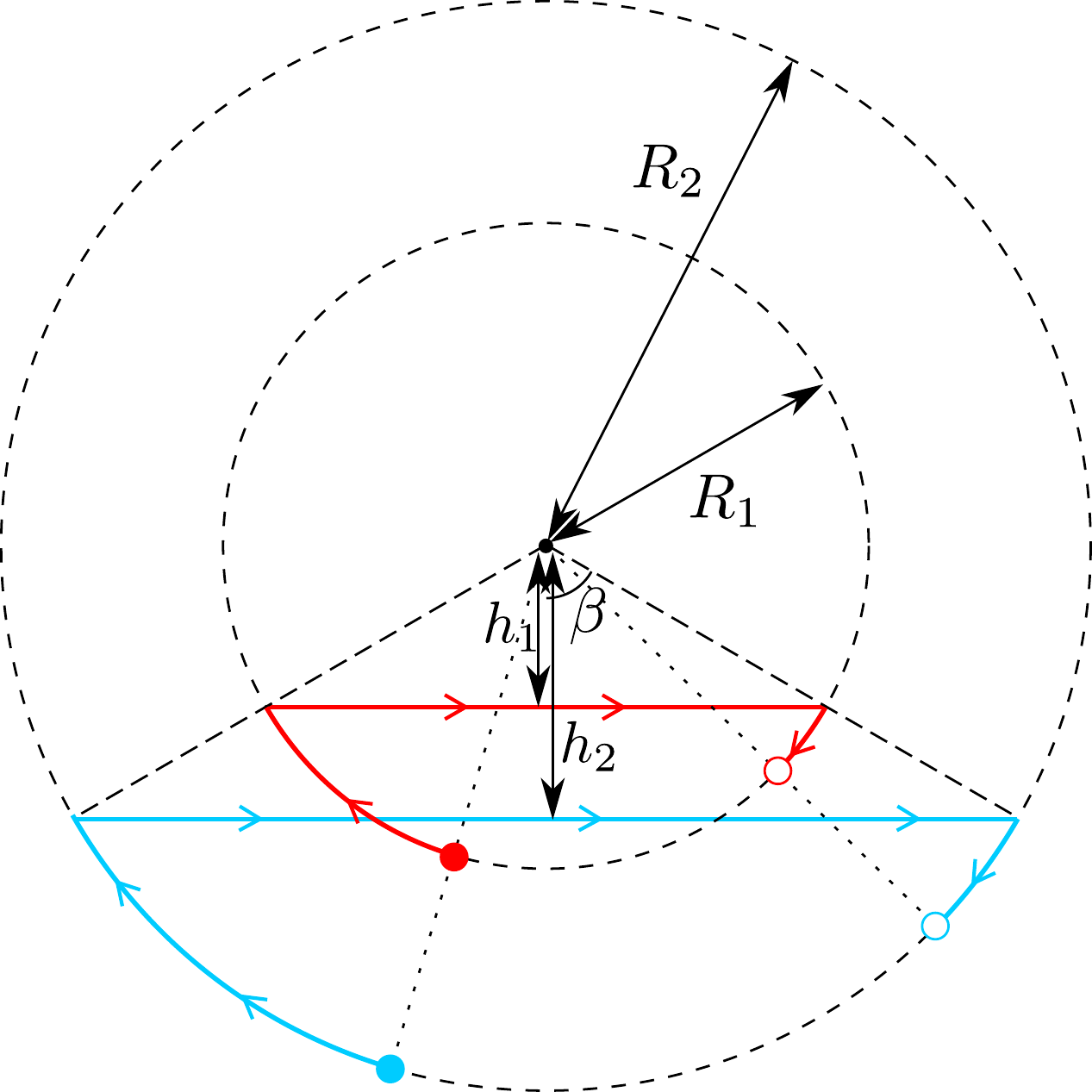}
\caption{Trajectories of two tracer particles (red, blue) in the plane $z=0$ under the $z$-axis rotation phase of the BST map (side view). The red tracer particle is on the inner spherical shell with radius $R_1$, with a fill depth $R_1 -h_1$; the blue particle is on the outer spherical shell with radius $R_2$, with fill depth $R_2 - h_2$ such that $h_1/R_1=h_2/R_2$. Blue and red particles start and finish on the same radial lines (dotted).}
\label{fig:LR_equiv_schematic}
\end{figure}

\begin{figure}[tbp]
\centering
\includegraphics[width=\columnwidth]{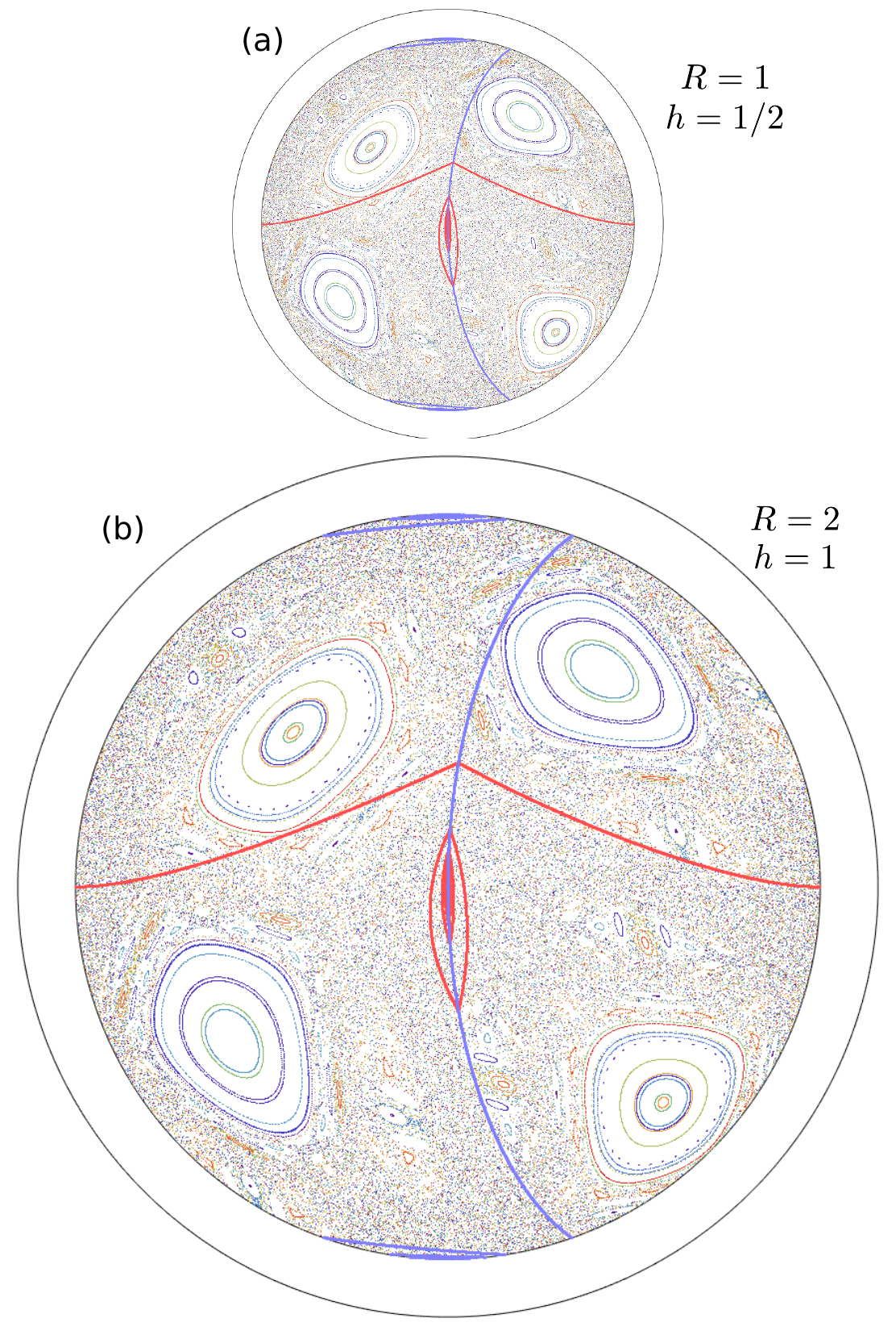}
\caption{Poincar\'{e} sections and cutting lines [thick blue and red (gray) curves] for the BST map with $\theta_z = \theta_x = \pi/3$ (bottom view). (a)~$h=0.5$, $R=1$. The outer circle has radius equal to $1$. (b)~$h=1$, $R=2$. The outer circle has radius equal to $2$.}
\label{fig:LR_equiv_psections}
\end{figure}

Considering the connection between the fill depth, $R - h$, and the spherical shell radius, $R$, Fig.~\ref{fig:LR_equiv_schematic} shows that for constant $\beta = \arccos(h/R)$, particles in the plane $z=0$ on the same initial radial line experiencing the same rotation about the origin, end up on the same radial line. This is because all lengths in the system are scaled equivalently. Therefore, the BST map with rotation angles $\theta_z,\theta_x$ on the spherical shell with radius $R_1$ and fill depth $R_1 - h_1$ will produce identical mixing and transport (up to scaling) to the map with radius $R_2$ and fill depth $R_2 - h_2$ as long as $h_1/R_1=h_2/R_2$. For instance, Fig.~\ref{fig:LR_equiv_psections} shows that the Poincar\'{e} sections for the $\theta_z=\theta_x=\pi/3$ rotation protocol are identical (up to a scaling factor) when $h/R=1/2$. Consequently, we are able to study the mixing and transport dynamics across the whole $(h,R)$ parameter space by keeping $R=1$ fixed and varying $h$ between $0$ (half full) and $1$ (empty).

%\bibliographystyle{apsrev4-1}
%\bibliography{mybib}

\end{document}